\documentclass[%
 reprint,
 amsmath,amssymb,
 aps,
]{revtex4-2}
\usepackage{appendix}
\usepackage{booktabs}
\usepackage{todonotes}
\usepackage{marginnote}
\usepackage{multirow}
\usepackage{xcolor}
\usepackage{caption}
\usepackage{subcaption}
\usepackage{graphicx}
\usepackage{dcolumn}
\usepackage{bm}

\newcommand{\beginsupplement}{%
        \setcounter{table}{0}
        \renewcommand{\thetable}{A\arabic{table}}%
        \setcounter{figure}{0}
        \renewcommand{\thefigure}{A\arabic{figure}}%
     }
\newcommand{\beginsuppinf}{%
        \setcounter{table}{0}
        \renewcommand{\thetable}{S\arabic{table}}%
        \setcounter{figure}{0}
        \renewcommand{\thefigure}{S\arabic{figure}}%
     }
\begin{document}

\preprint{APS/123-QED}

\title{New geometric receipts for design of photonic crystals and metamaterials: optimal toric packings}

\author{ A.Itin}
    \email[Email address and website: ]{alx.itin@gmail.com, AI4science.de}
    \affiliation{TUHH, Hamburg, Germany}

\date{\today} 

\begin{abstract}
Design of photonic crystals having large bandgaps above a prescribed band is a well-known physical problem with many applications. A connection to an interesting mathematical construction was pointed out some time ago: it had been conjectured that optimal structures for gaps between bands n and n+1 correspond, in case of transverse magnetic polarisation, to rods located at the generators of centroidal Voronoi tessellation (CVT), and in case of transverse electric polarisation to the walls of this tessellation. We discover another mathematical receipt which produces even better solutions: optimal packing of discs in square and triangular tori. It provides solutions qualitatively different from CVT, sometimes increasing the resulting bandgap size in several times. We therefore introduce two new classes of periodic structures with remarkable properties which may find applications in many other areas of modern solid state physics: arrays of particles located at the centers of optimally packed discs on tori, and nets corresponding to the walls of their Voronoi tessellations. 

\end{abstract}
\maketitle

Photonic bandgaps in  periodic, aperiodic, and quasicrystal materials have been studied intensively during the last decades \cite{Yablon87, Yablon93, John, Johnson, GeometricOptimal, Preble, EdagawaReview, Lucas, Christensen, Glotzer, Dyachenko, Dyachenko2, topMen, SFan, Man, Torquato,Matthews, Rechtsman, foam, foam2,foam3}.
These topics have important technological applications for control and manipulation of light and are also encountered in nature \cite{Nature}.
Usually design of photonic structures is fulfilled numerically using various optimisation methods, or even machine learning techniques, or their combinations. For these approaches, good initial guesses would be very helpful. Optimality of the final designs are not always known: combinatorial complexity of the design space is so huge that one usually cannot be sure whether to continue search for the optimal solution, or the limit of effectiveness is (nearly) reached. Therefore despite advances in numerical and machine learning techniques \cite{ML2,ML3,ML4}, understanding design principles of photonic bandgap structures would be beneficial for current and future technological progress, and may be useful for more advanced photonic problems (e.g. \cite{metaphotonics,metaphotonics2}) and many adjacent areas of scientific knowledge (e.g. plasmonics \cite{plasmonRev,plasmon1,plasmon2}, phononics \cite{phononRev,phonon1,phonon2,phonon3}, phoxonics (optomechanics) \cite{phoxonics,phoxonics2}, and magnonics \cite{magnon1, magnonRev}). Furthermore, numerical approaches generally lacks interpretability: in case the final design is very efficient, what geometric features determine its effectiveness?
An interesting mathematical receipt for photonics crystals design was suggested in \cite{GeometricOptimal} and was shown to be superior to previous techniques.  It contains two steps. The first one is purely geometric:  finding {\em centroidal Voronoi tessellation} (centroidal VT (CVT) ) of the unit cell on $n$ domains. This is a geometric construction where $n$ generating points are placed on a plane in such a way that after building VT each point coincides with a center of mass of its Voronoi cell. In practice, this is achieved numerically using Lloyd algorithm.  Putting spots of an optical material at locations of the generators of CVT  would result in a bandgap above the n-th band in the banddiagram. In the second step, topological optimisation of the initial spots is fulfilled (which involves numerical optimisation in the space of hundreds of variables, as the state of each pixel is a variable (design parameter)).  

\begin{figure}[hbt]
\caption{Optimal packings of discs on a 2D square torus. From top left to bottom right: N =2 to N=10 (the last configuration N=10 is a conjectured optimal packing from Ref.\cite{Connelly2}). Each figure depicts unfolding of a torus on a plane. 2x2 supercell is shown for clarity: the torus corresponds to 1/4 of each picture.  \label{fig:sqtoripack} }
\centering
\includegraphics[width=0.48\textwidth]{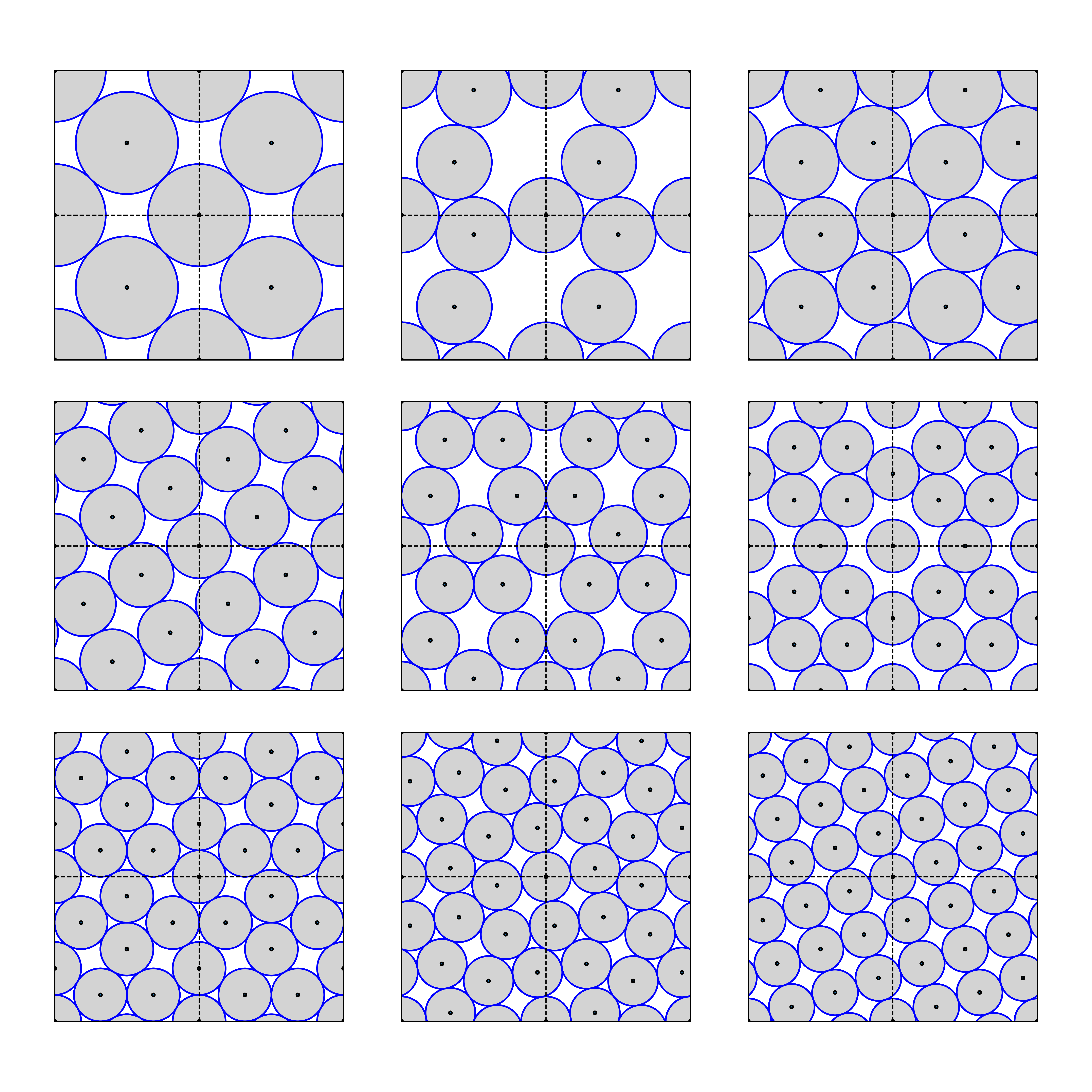}
\end{figure}
\begin{figure}[h]
\caption{Optimal packings of discs on a 2D triangular torus. From top left to bottom right: N =2,3,4,5. Each picture contains 4 unfoldings of the torus. N=4 is a "trivial" perfect packing, obtained by 2x2 "supercelling" of N=1 case. N=3 case is nontrivial. \label{fig:trtoruspack} }
\centering
\includegraphics[width=0.45\textwidth]{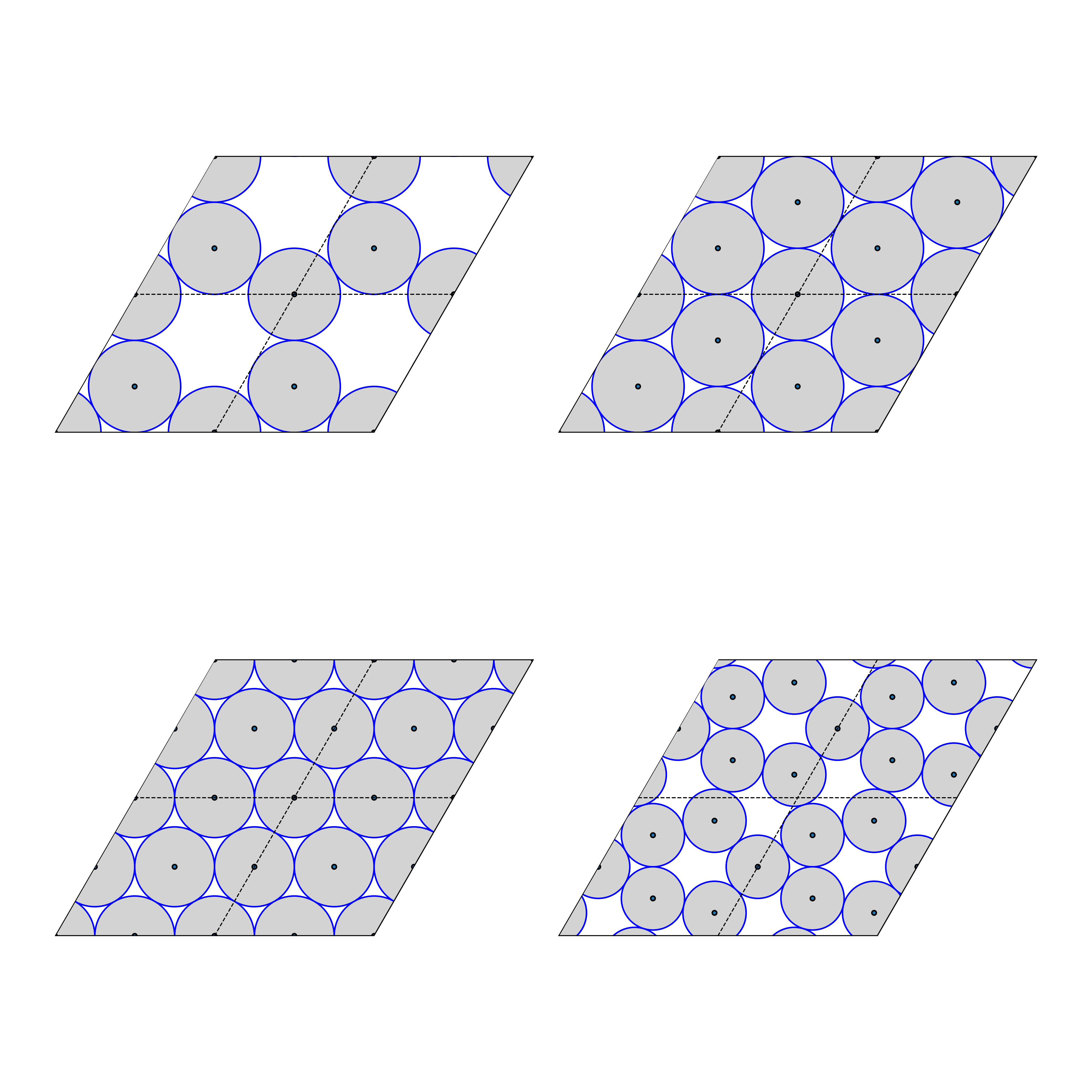}
\end{figure}

\begin{figure*}[t]
\caption{Relative bandgap vs. band number for square lattice geometry (left panel) and triangular lattice (right panel). Upper(bottom) panel: TM(TE) modes. Dot-dashed line ("Lloyd+rods(wires): CVT$_1$"): configurations realising centroidal VT (CVT) are taken from Ref. \cite{GeometricOptimal}, and circular rods of equal radius are placed at each point of the configuration for TM modes, or wires of equal width are placed along walls of VT for TE modes. The rod radius (wire width) is optimised to achieve the highest bandgap (1-parameter optimisation). Dashed line ("Lloyd+TO: CVT$_{top}$"): results of Ref.\cite{GeometricOptimal}, where TO was done on top of CVT. Solid line ("Opt.packing+ rods(wires): OP$_1$" for TM(TE) modes): our results, where configurations are obtained from optimal packing solutions and circular rods of equal and optimised radius are placed there for TM modes, or wires of optimised width are placed along walls of VT for TE modes. Stars denote perfect packing solutions (close-packed structures), those VT in TE case produce honeycomb-like structures with maximal bandgaps. They are realised for "magic" band numbers $N=1,4,9$ found both in \cite{GeometricOptimal} and in our approach and at additional "magic" band numbers $N=3,7,12,13$ obtained via our approach only (highlighted by vertical arrows). 
(Color online).\label{fig:bandgaps}  }
\centering
\begin{subfigure}[]{0.45\textwidth} 
\subcaption{TM, square}
\includegraphics[width=\textwidth]{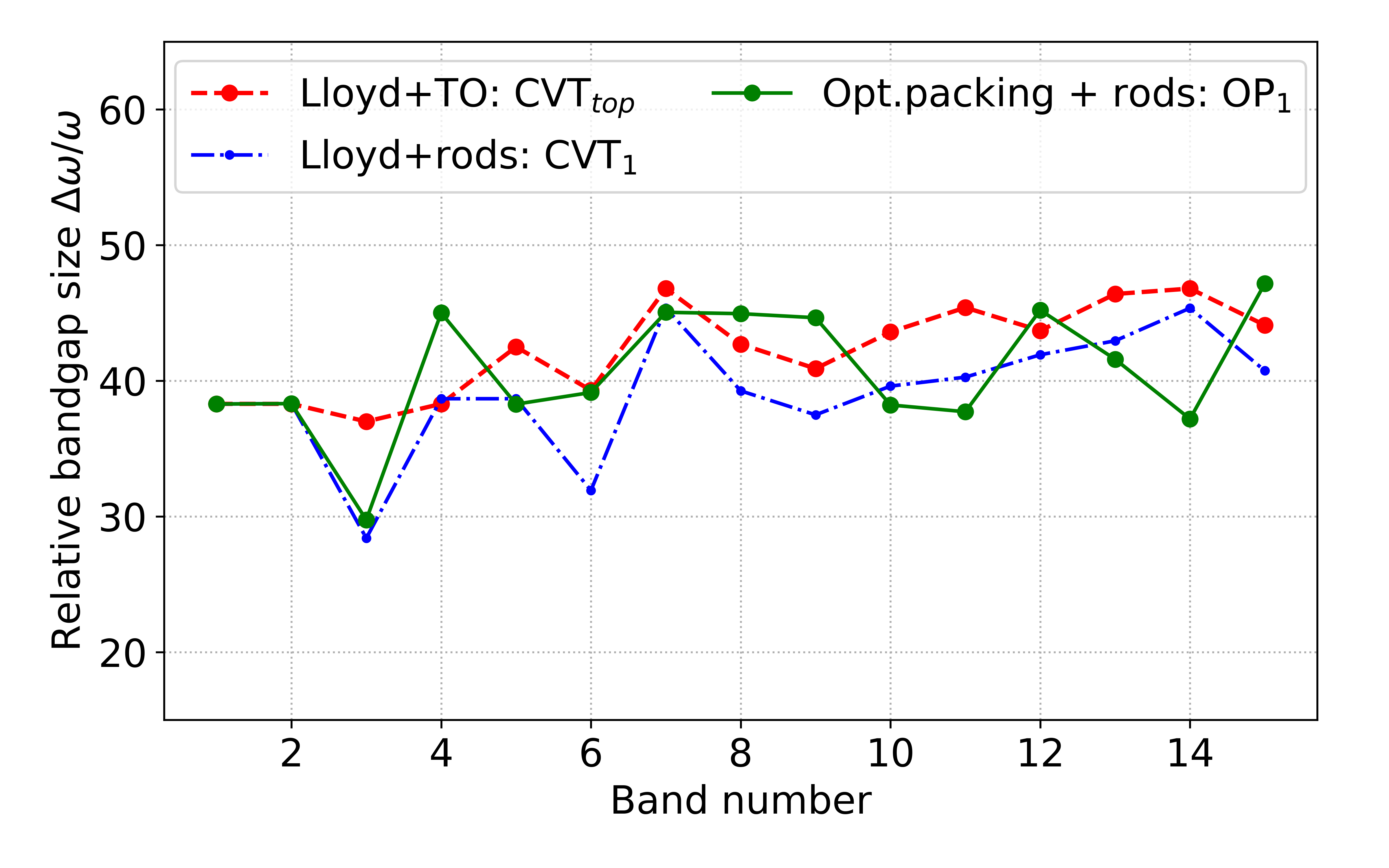}

\end{subfigure}
\begin{subfigure}[]{0.45\textwidth}
\subcaption{TM, triangular}
\includegraphics[width=\textwidth]{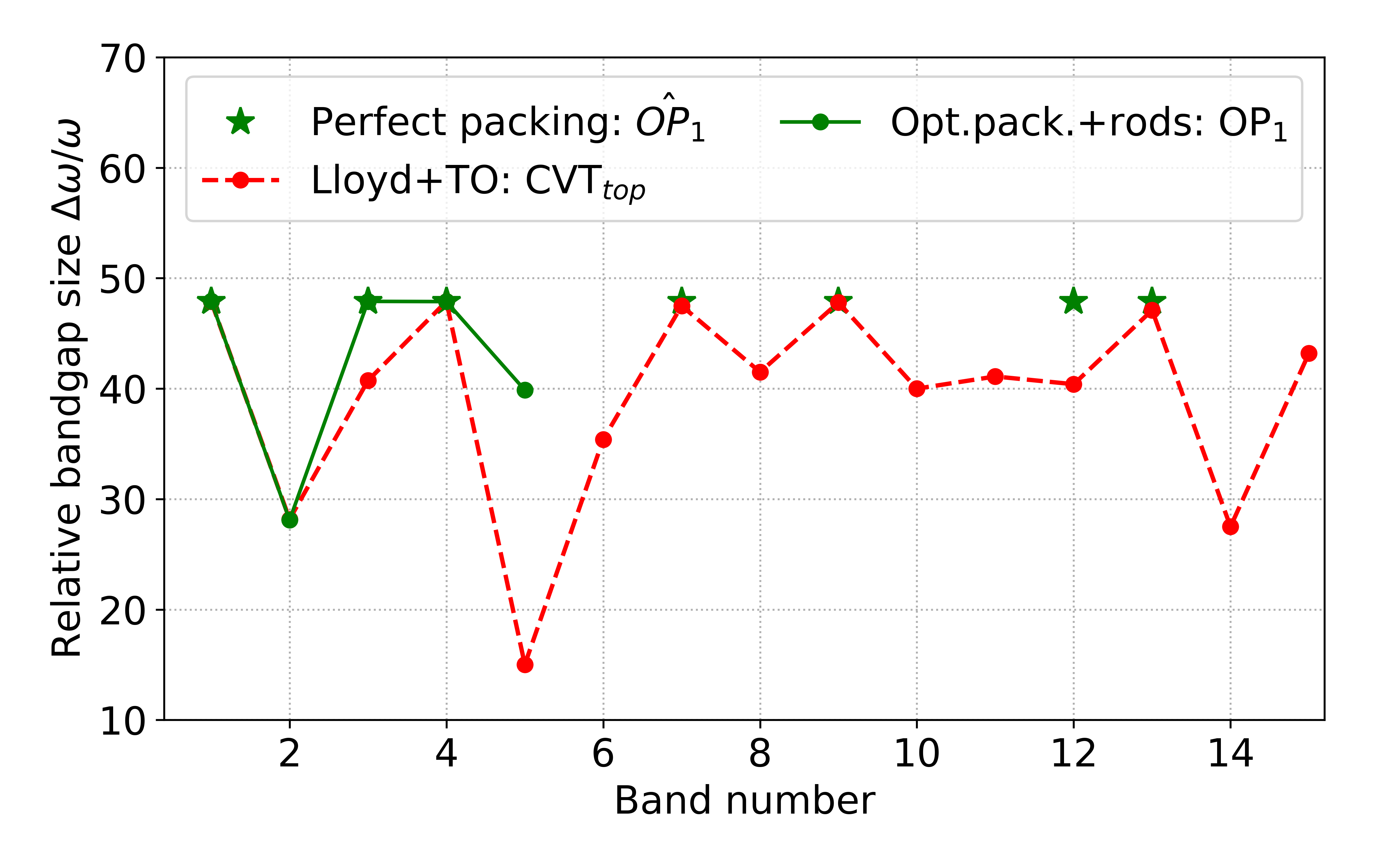}
\end{subfigure}
\begin{subfigure}[]{0.45\textwidth}
\subcaption{TE, square}
\includegraphics[width=\textwidth]{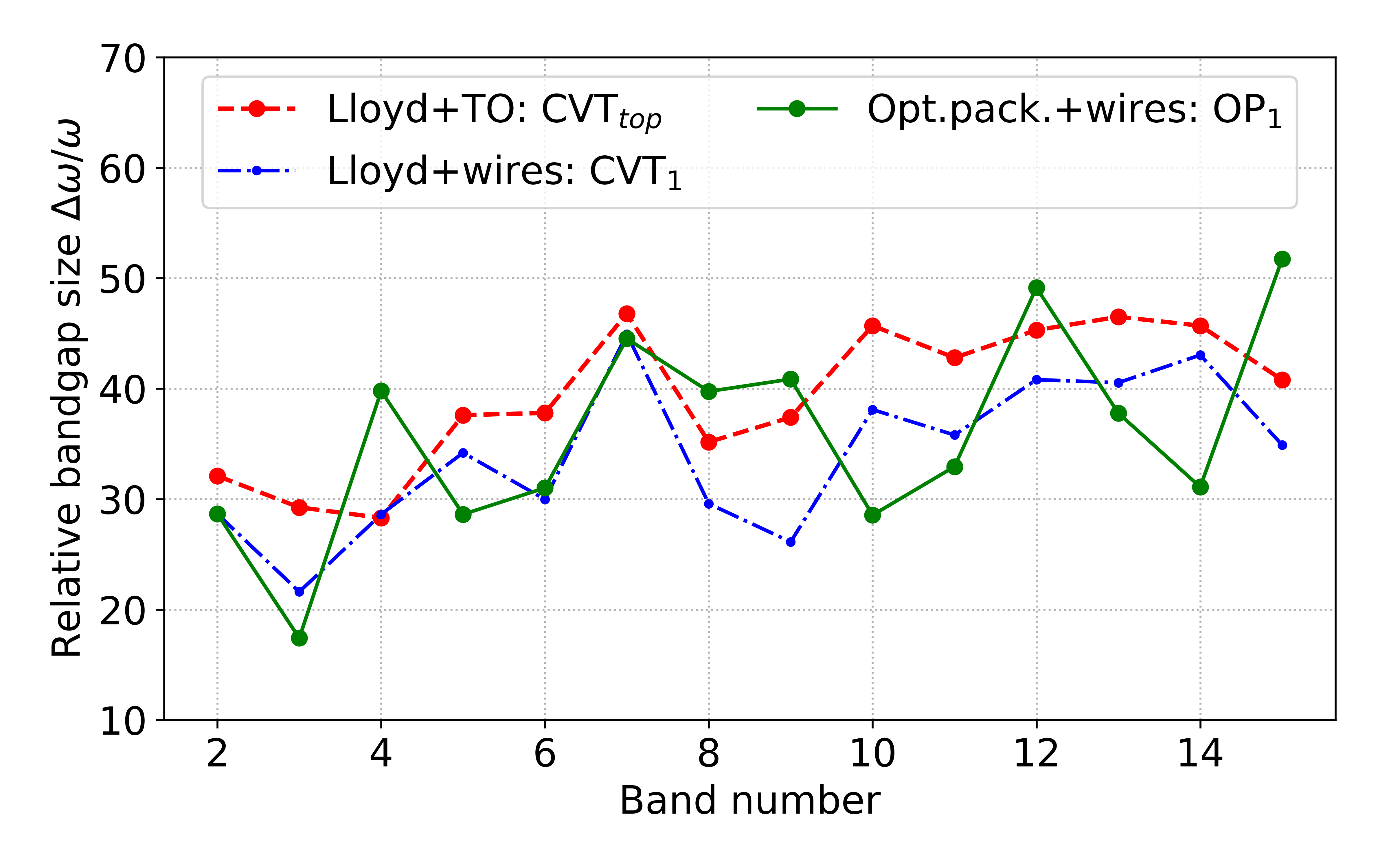}
\end{subfigure}
\begin{subfigure}[]{0.45\textwidth}
\subcaption{TE, triangular}
\includegraphics[width=\textwidth]{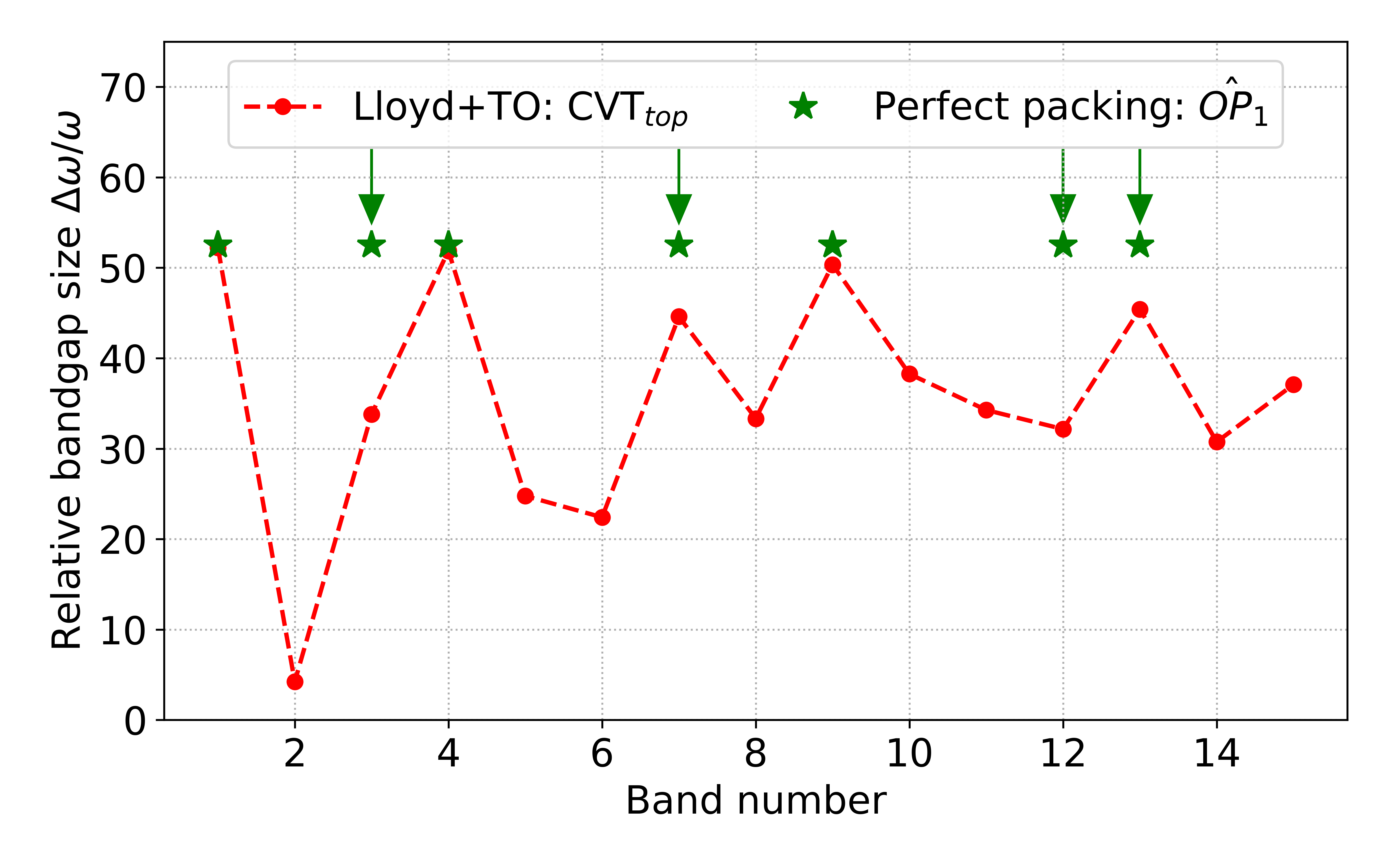}
\end{subfigure}
\end{figure*}

Here we suggest a different geometric receipt and new classes of structures that considerably improve results of \cite{GeometricOptimal} and possess remarkable properties which may be beneficial in other fields as well.


The problem formulation with respect to photonic bandgaps is as follows \cite{GeometricOptimal}:
consider distribution of dielectric materials which is periodic in the $xy$-plane and constant in the $z$ direction, i.e. $\varepsilon(\bf{r} + \bf{R_j}) = \varepsilon(\bf{r})$, where $R_j$ are lattice basis vectors with zero $z$ component.  For simplicity we also assume here that the dielectric function $\varepsilon$ is piecewise-constant and only attains one of the 2 values: $\varepsilon_1$ or $\varepsilon_2$. 
The Maxwell equations for electromagnetic waves propagating in the xy-plane are decoupled on two wave equations for TM (E-field in the z- direction) and TE (H-field in the z- direction) polarized waves:
\begin{eqnarray}
\mbox{TM} &:& \nabla^2 E_z ({\bf{x}}) + \frac{\omega^2 \epsilon({\bf{x})}}{c^2} E_z ({\bf x}) =0, \label{TM} \\
\mbox{TE} &:& \nabla \cdot \Bigl[ \frac{1}{\epsilon ({\bf x}) } \nabla H_z ({ \bf x}) \Bigr] \frac{\omega^2 }{c^2} H_z ({\bf x}) =0. \label{TE}
\end{eqnarray} 

The scalar fields satisfy the Floquet-Bloch wave conditions $E_z = e^{i {\bf k \cdot x}} E_k$ and $H_z = e^{i {\bf k \cdot x}}H_k$, respectively, where $E_k$ and $H_k$ are cell periodic fields \cite{GeometricOptimal}. 
Solving e.g. Eq.\ref{TM} for wavevectors $k$ within the first Brillouin zone, one gets banddiagram (a set of bands $\omega_n(k)$ ) which posesses a bandgap between bands $n$ ans $n+1$ if  $\min{\omega_{n+1}(k)} > \max{\omega_n(k)}$.
The question is, how to design the periodic distribution of dielectric  that maximizes the relative band gap above any particular band n?

Before outlining another geometric approach, let us think about the meaning of the procedure of Ref.\cite{GeometricOptimal}. 
Intuitively, CVT aims at placing generating points somehow uniformly. In case a generating point happens to be at the geometric center of mass of its Voronoi cell, it implies certain level of "uniformity" of the points locations. However there are many different ways to define "uniformity". What if we choose a simpler and hence probably a more "fundamental" principle: maximization of the minimal distance between the points?

\begin{figure*}[hb!t]
\caption{Walls of Voronoi tesselations (VT) of optimal packing configurations (a) vs walls of centroidal VT obtained in \cite{GeometricOptimal} (b). Each panel shows cases from N=2 (top left) to N=10 (bottom right). Cases up to N=15 see in Appendix. The points generating VT are centers of optimally packed discs in (a) or generators of CVT in (b).
\label{fig:VT}}
\begin{subfigure}[]{0.48\textwidth}
\subcaption{Optimal packings and VT}
\includegraphics[width=\textwidth]{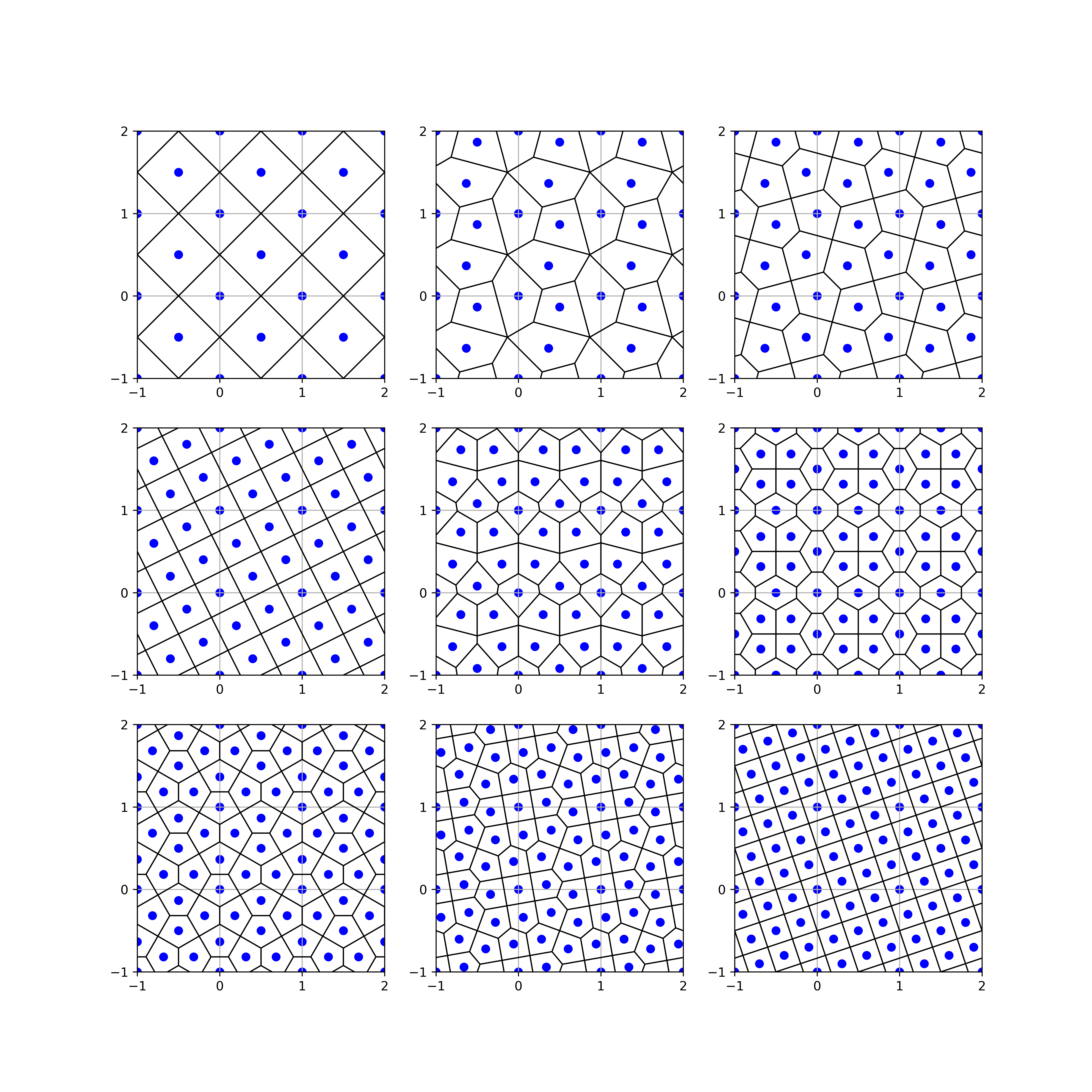}
\end{subfigure}
\begin{subfigure}[]{0.48\textwidth}
\subcaption{Centroidal VT}
\includegraphics[width=\textwidth]{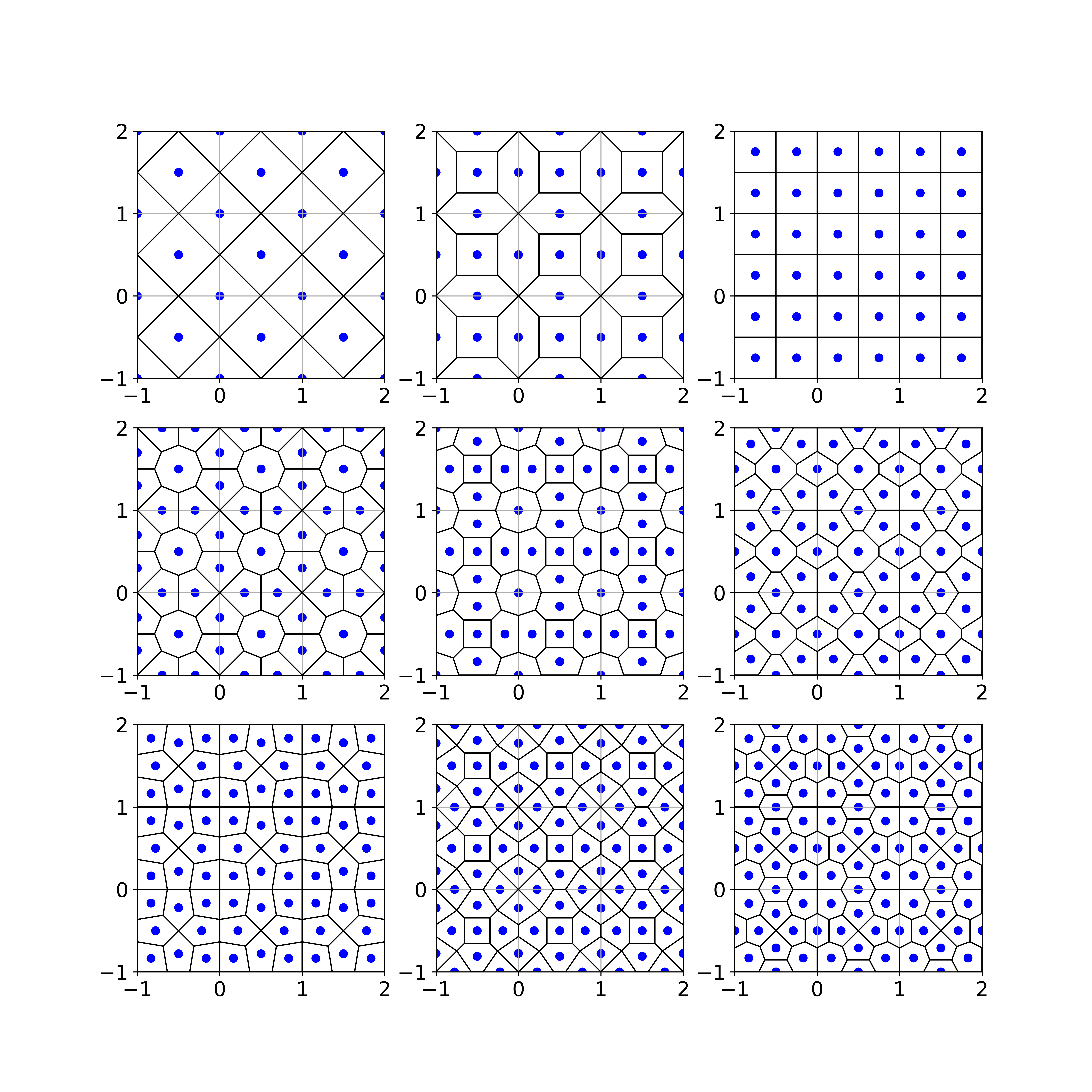}
\end{subfigure}
\end{figure*}

Actually, in that case we obtain a set of complicated mathematical problems that were not solved yet at the time of creation of insightful Ref.\cite{GeometricOptimal}!
Nowadays we can use modern results obtained in the area of mathematics dealing with geometric packing problems. 
Packing problems have arisen throughout the history of humankind \cite{Packing}, one of the most famous example is probably the Kepler problem. Kepler, who "sought order and structure within solid matter, just as he did in the solar system" \cite{Packing2}, was thinking about densiest arrangement of (many) identical spheres, conjecturing a close-packed structure as a solution. It took 300 years to prove 2D version of the Kepler conjecture, and another century to prove much more difficult 3D case. These results are about infinite space. We would need somehow related results for finite space: optimal packings in "containers" with periodic boundary conditions. Indeed, eventhough a photonic crystal occupies infinite space, it is obtained by periodic arrangement of a finite elementary unit cell. We would like to distribute n points within this unit cell. Optimal packing of n discs which maximise minimal distance between n {\em centers} of these discs is exactly what we need. Therefore, the mathematical concept we are going to use is optimal packings of discs in a square {\em torus} (for a square unit cell) and a triangular {\em torus} (for rhombic unit cell).  On Figs.\ref{fig:sqtoripack},\ref{fig:trtoruspack}, and \ref{fig:sqpack} (see \cite{SuppInf}) we show optimal packings of discs in a square torus, a triangular torus, and a square, correspondingly. These are different problems with different solutions. Moreover, unlike Kepler conjecture, not everything is known for these problems yet. The last configuration N=10 of Fig.\ref{fig:sqtoripack} is only a conjecture. For a square torus, optimal packings of up to N=5 discs were shown in Ref. \cite{Dickinson}, and for N=6,7,8 by Musin and Nikitenko in Ref.\cite{Musin12}, who also conjectured N=9 optimal configuration. This conjecture was confirmed by Smirnov and Voloshinov in Ref.\cite{Smirnov} using reformulation to a mixed-integer nonlinear problem. This is where current mathematical frontier is located (it is not known yet how to place  N=10 points on a square torus optimally), but for our purposes we need configurations of up to N=15 points. We take them (from N=10 to 15) from the conjectured optimal configurations of Ref.\cite{Connelly2} (see Figs.\ref{fig:sqtoripack15} in Appendix). For a triangular torus, we use optimal configurations of up to N=5 from Ref. \cite{Dickinson2}, and also a very useful result on perfect packings from Ref. \cite{Connelly} to be discussed below. We do not use topological optimisation (TO) because it would actually make difficult the comparison with Ref. \cite{GeometricOptimal} (since there could be many variants of TO). Our focus is on the geometric step, and after fulfilling it, we make only simple 1-parameter optimisation, varying a single parameter: radius of rods (in TM case), or width of wires in nets (in TE case) to achieve a larger bandgap. If 1-parameter optimisation is comparable with or better than TO (which involves hundreds of parameters), it is a clear sign of success.   
Results for TM polarisation in square lattice geometry is shown in Fig. \ref{fig:bandgaps}a. We apply 1-parameter optimisation to our configurations and also to CVT configurations of \cite{GeometricOptimal}  by putting circular rods of equal radius at configuration points and varying the radius to increase the bandgap. Let us denote these approaches OP$_1$ and CVT$_1$, correspondingly. Results of full topological optimisation of Ref.\cite{GeometricOptimal} are also shown, let us denote it CVT$_{top}$ ("Lloyd+TO:CVT$_{top}$" on Fig. \ref{fig:bandgaps}a).  Permittivity of the rods is the same as in Ref.\cite{GeometricOptimal}: $\varepsilon= 11.56$, while host media is air with $\varepsilon=1$. Up to $N=9$ our OP$_1$ perform better or the same as CVT$_1$. From $N=10$ the results alternate.  Surprisingly, at 4 cases out of 15 (N=4,8,9,15) OP$_1$ overcomes even CVT$_{top}$ and at 5 cases they are equal or approximately equal. Especially $N=4$ case is interesting. In CVT case the points form a simple square lattice and boundaries of VT form a square net (see Fig. \ref{fig:VT}b, top right panel). In OP case quadrumers of rods are shifted along each other forming a peculiar configuration, and boundaries of VT form a net consisting of pentagons (see Fig. \ref{fig:VT}a, also top right panel). This leads to considerable enlargement of the bandgap both in TM and TE cases. In TE case (Fig. \ref{fig:bandgaps}c for square geometry),  we also vary a single parameter: width of the walls of nets lying along boundaries of our VT (also denoting this approach OP$_1$) and boundaries of CVT (again, CVT$_1$).  
In N=8 case, optimal packing configuration can be obtained from N=4 case by rescaling and rotation on 45°. Therefore, optimised bandgap values for N=4 and N=8 cases coincide.
In N=7 case, CVT of Ref.\cite{GeometricOptimal} and OP configuration coincide. This structure is remarkable also for its property to sustain a band gap even at the low permittivity values (i.e. at low contrast),  as  discussed in Appendix.
Important feature of TE square lattice case (Fig.\ref{fig:bandgaps}c) is inability of CVT$_{top}$ to achieve global maximum of the TE triangular lattice case. Note that in the triangular TE case (Fig.\ref{fig:bandgaps}d) CVT$_{top}$ 
achieves  maximal bandgap values of about 52$\%$ at "magic" band numbers N=1,4,9: a net forming perfect honeycomb network appears in these cases. In a square lattice, such values are not achievable for CVT$_{top}$, at least for the first N=15 bands. However, for OP$_1$ this result is possible! OP$_1$ overcomes CVT$_{top}$ at 5 cases (N=4,8,9,12,15), and, most importantly, achieves or overcome value of 50\% at $N=12,15$. Structures for N from 10 to 15 are shown in Appendix. OP seem to produce approximants of honeycomb lattice better than CVT.
N=4 case mentioned above produces peculiar pentagonal teselations in TE case. Such net is called Cairo tesselation \cite{CairoMorgan, Cairo, CairoConway} and is encountered in architecture  \cite{CairoMorgan}, decorative design \cite{Cairo}, and crystallography \cite{CairoCryst,CairoNat}, but was not used in photonic crystals yet, to the best of our knowledge.

Let us now consider the triangular lattice (i.e. rhombic unit cell) case in a more detail. In the TM case, we fulfill calculations within OP$_1$ approach for N up to 5. For all cases, OP$_1$ is the same or higher than CVT$_{top}$ (so we do not even need CVT$_1$ for comparison). At N=5 OP$_1$ bandgap is several times higher than in CVT$_{top}$ approach (40\% against approx. 15\%). Note a very important feature:  OP$_1$ achieves CVT$_{top}$ maximal value of approx. 48\% at N=1,3,4,..,  while CVT$_{top}$ at N=3 is considerably smaller. In TE case (Fig.\ref{fig:bandgaps}d),  CVT$_{top}$ achieves maximal bandgap at N=1,4,9. This is related to the fact that at N=1 generators of CVT form triangular lattice (are located at the centers of discs forming close-packed structure), whereas walls of CVT form perfect honeycomb network (see Ref.\cite{GeometricOptimal}). Clearly, close-packed structure and corresponding honeycomb network appears also at "supercells" of N=1 configuration, i.e. N=4 (2x2 supercell), N=9 (3x3), and so on. In other words, for N=$a^2$, where $a$ is an integer. 
However, N=$a^2$ are not the only possible "magic" band numbers! Looking at OP configuration at N=3 (Fig. \ref{fig:trtoruspack}), we also see perfect close-packed structure (let us call it perfect packing for short).  
Actually, for the triangular torus maximal packing density of discs (perfect packing) can be achieved if the number N is of the form $N = a^2 + ab + b^2$, for $a, b$ integers \cite{Connelly}. These numbers are called Loeschian numbers in \cite{Sloan}, and triangular lattice numbers in \cite{Connelly}.  They form Sloane sequence A003136 \cite{Sloane}. The first few nonzero numbers of this form are  1, 3, 4, 7, 9, 12, 13, 16. It means for the first 15 bands we can get maximal bandgap of around 52$\%$ for the "magic band numbers" N=1,3,4,7,9,12,13 instead of just N=1,4,9 as in the approach of \cite{GeometricOptimal}. We denote 1-parameter optimisation over such configurations $\hat{OP_1}$ in Figs. \ref{fig:bandgaps}b,d and only show $\hat{OP_1}$ and CVT$_{top}$ in Fig. \ref{fig:bandgaps}d to emphasize such peculiar finding. Interestingly, such perfect packings simultaneously fulfill CVT condition. But if we postulate CVT approach as our geometric receipt, we are left with the numerical Lloyd algorithm which can produce various outcomes (and finds no nontrivial "magic band numbers", i.e. different from $N^2=a^2$, in Ref.\cite{GeometricOptimal}). If we postulate optimal packing as our geometric receipt, we immediately have enlightening support from powerful mathematical results of optimal packing research, giving us in particular triangular lattice numbers as "magic band numbers". 
To conclude, we established a connection between area of optimal packing problems and optimised photonic crystals. Since the geometric steps we discussed do not depend on Maxwell equations, the obtained structures may have applications in various fields besides photonics.  The results might increase activity not only in the fields of solid state physics, but in optimisation theory as well. Modern computers, classical or quantum, allow solutions of packing problems beyound $N=9$ (which required about 20.000 CPU hours in Ref.\cite{Smirnov}). Probably absence of clear physical applications diminish activity in this fundamental area of research. Here we provided examples of such applications. The results may lead to new technological applications. In particular, using Nanoscribe’s two-photon polymerization technology \cite{nanoscribe}, as well as other techniques, it is possible to create polymer-based photonic structures. There, usually two problems arise: the structures should be simple enough (for robustness against mechanical deformations and manufacturing errors), and posess a bandgap at low contrast.  As shown in Appendix, the best structures discussed here sustain bandgaps up to very low contrasts which makes them suitable for wide range of materials and technologies.  The last but not least, our work provides a surprising connection between art and science, e.g. by finding a decorative ornament (Ref.\cite{Cairo, CairoMorgan}) which upon 1-parameter optimisation outperforms fully topologically optimised CVT$_{top}$ photonic crystal for N=4.

\section*{Acknowledgments}
We thank organisers of NANOP23 \cite{NANOP23} for invitation to that exciting and insightful conference, M.Berry for inspirational discussions there, German Research Foundation (DFG) for supporting that visit via SFB 986 “Tailor-made Multi-scale Materials Systems: M3”, Project 192346071S, A.Petrov and M.Eich for the invitation to TUHH, discussions, and for bringing attention to Ref. \cite{Lucas}, S. Smirnov and A.Tarasov for sharing details of their sophisticated calculations.  

\clearpage

\appendix
\beginsupplement
\section{ End matter }
\begin{figure}[hbt]
\caption{Conjectured optimal packings of discs on 2D square torus. From top left to bottom right: N =10,11,12,13,14,15. Each figure depicts periodic unfolding of a torus on a plane. 2x2 supercell is shown for clarity, i.e a torus correspond to 1/4 of each picture. \label{fig:sqtoripack15} }
\centering
\includegraphics[width=0.5\textwidth]{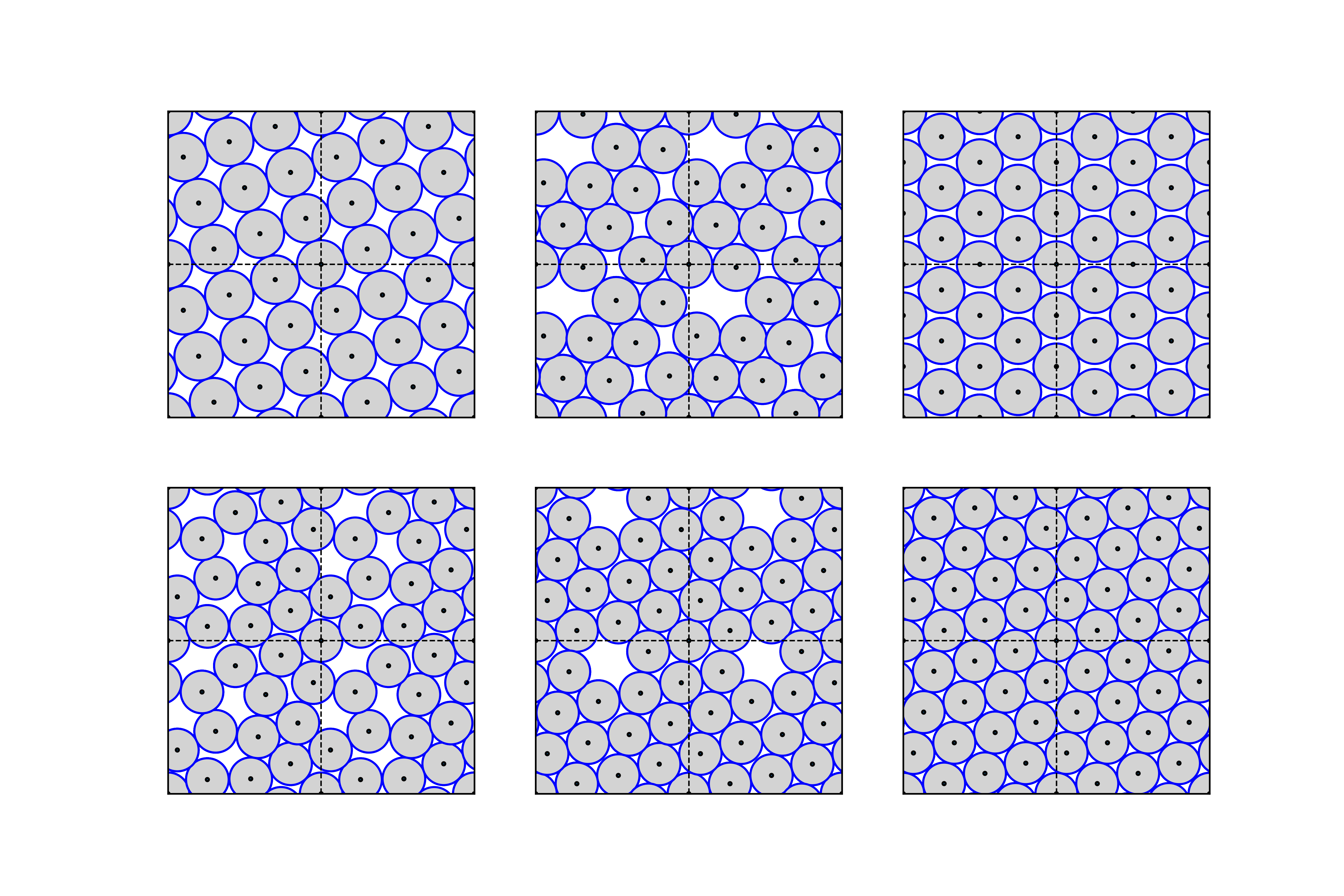}
\end{figure}

Conjectured optimal packing configurations of discs on 2D square torus for N=10-15 discs are available at Ref.\cite{Connelly2} and are shown on Fig. \ref{fig:sqtoripack15}. Mathematical estimates of \cite{Connelly2} confirm that these configurations are close to the optimal, but whether they are optimal or not is not known. A question arise, should we try to search for even better configurations? In Fig. \ref{fig:bandgaps}a of the main text it can be seen that up to N=9 OP$_1$ approach for TM modes gives better or the same results as CVT$_1$  (both are 1-parameter optimisations, where rods are placed either at points of CVT configurations, or optimal packing configurations). However at N=10 CVT$_1$ is slightly better. Actually, it can be checked that OP configuration at N=10  is "better" than CVT in the sense of minimal distance between rod centers. But in CVT there are only few "defects" which spoil the minimal distance metric. Apart from these "defect" locations, minimal distance between other rods is greater than in OP configuration. This shows that as N is increased, situation becomes more complex and small number of "defects" becomes acceptable for the system. This question is considered elsewhere \cite{inprep}. Yet conjectured OP configurations continue to remain very good from the point of bangaps: at N=12 and N=15 they again overcome CVT$_1$ and even CVT$_{top}$.   

In Table \ref{tab:sqtable} we list parameters of optimal and conjectured optimal packings in square torus.
The parameters are: the geometric parameter $d$ (minimal distance between the points \cite{Musin12}), the optimised radius $r$ of rods for OP$_1$ configurations in TM case and the optimised width of wires $w$ for OP$_1$ configurations in TE case. Calculations are done with parameters mentioned in the main text (permittivity $\varepsilon$ of inclusions (rods or wires) is equal to $\varepsilon$=11.56, an permittivity of the host media is 1) using the package mpb \cite{mpb} on a grid 128x128.

\begin{figure}[hbt]
\caption{ CVT configurations of Ref.\cite{GeometricOptimal} from N=10 (top left) to N=15 (bottom right). 3x3 supercell is shown. Numbers just highlight individual unit cells. \label{fig:CVT15} }
\includegraphics[width=0.49\textwidth]{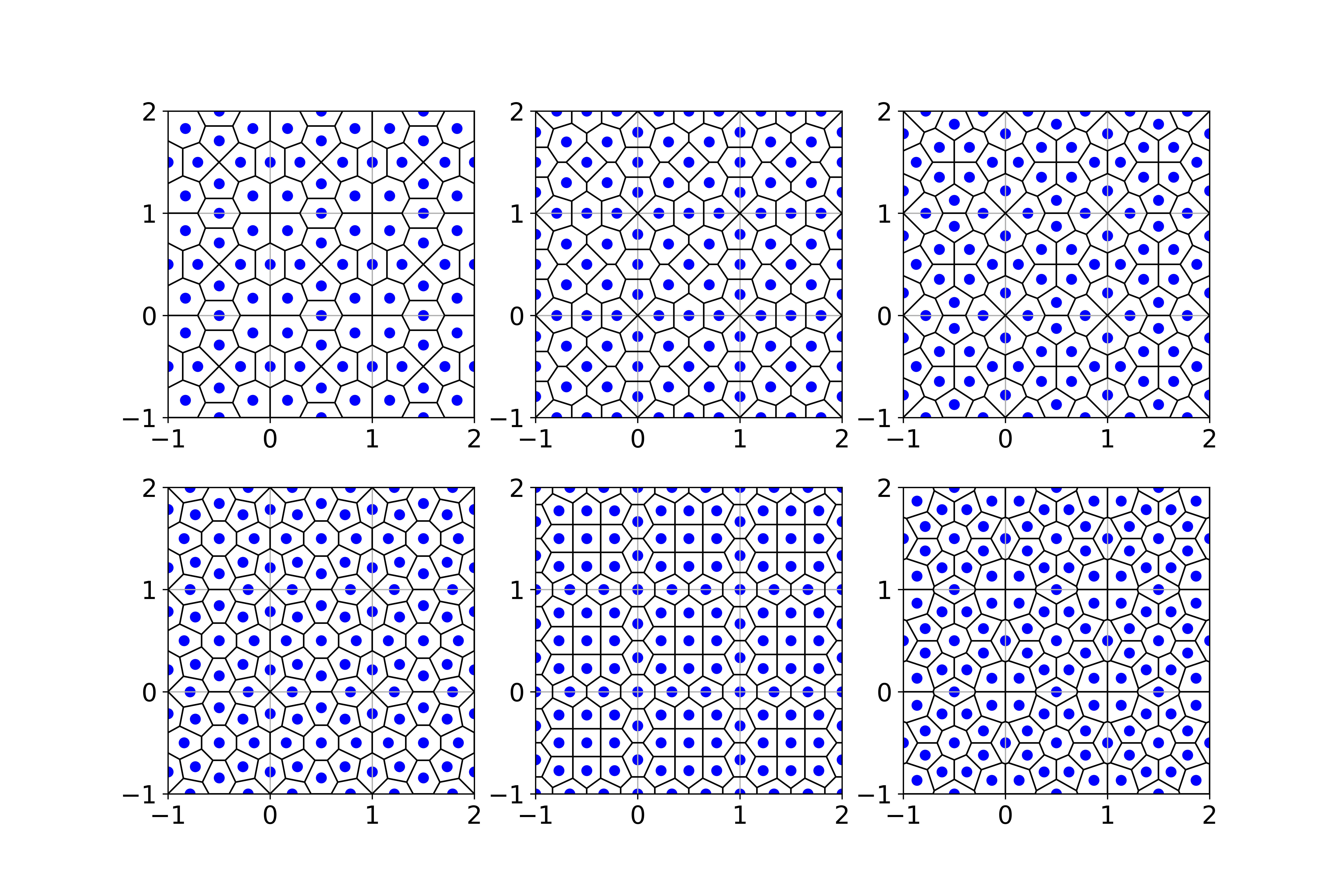}

\end{figure}

\begin{figure}
\caption{ VT of conjectured optimal packings in square tori.N=10 (top left) to N=15 (bottom right). N=12 and N=15 form structures that are very close to hexagonal lattices. \label{fig:OP15} }
\includegraphics[width=0.48\textwidth]{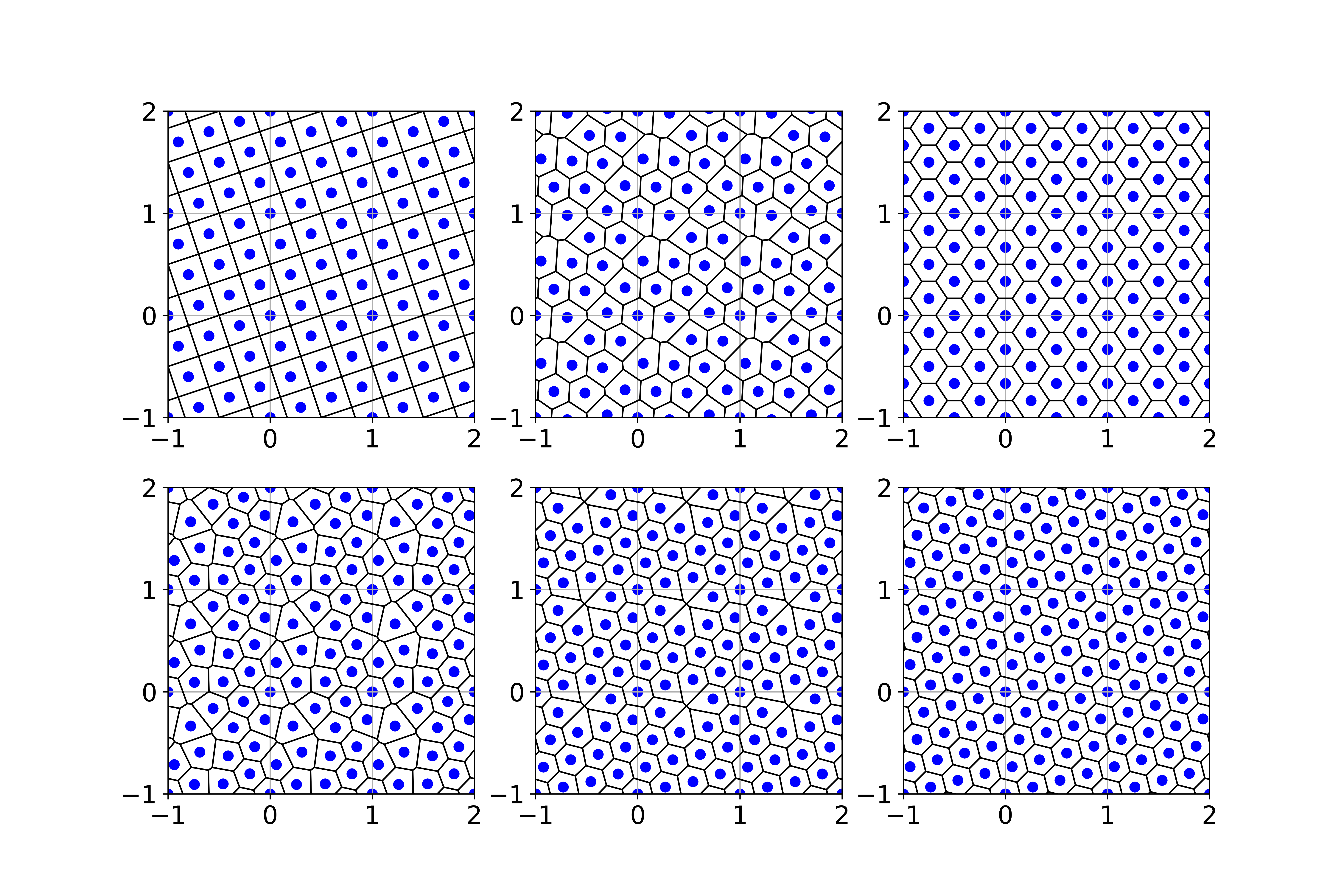}
\end{figure}

On Fig. \ref{fig:CVT15} we show CVT configurations of Ref. \cite{GeometricOptimal}. It is "continuation" of the Fig.\ref{fig:VT}b of the main text for $N=$10 to 15. For creating these figures, coordinates of points generating CVT were deduced (digitally extracted) from Figs. of Ref.\cite{GeometricOptimal} and VT were constructed using scipy Python package.
On Fig. \ref{fig:OP15} we show VT of conjectured optimal packing configurations. Generators of VT are centers of discs of conjectured optimal configurations of Ref.\cite{Connelly2} for N=10 to 15. Configurations for N=12 and N=15 form structures that are very close to honeycomb network structures. Configuration for N=14 is very interesting and resembles a variation of 6-fold floret pentagonal tiling \cite{CairoConway}. Due to different symmetry of the square unit cell it is not possible to produce such structure exactly, but the resulting figure clearly approximates it. 

\begin{figure}
\caption{ Optimised TE structure (honeycomb network) for triangular geometry for N=1. 3x3 supercell is shown for clarity.   \label{fig:hex} }
\includegraphics[width=0.4\textwidth]{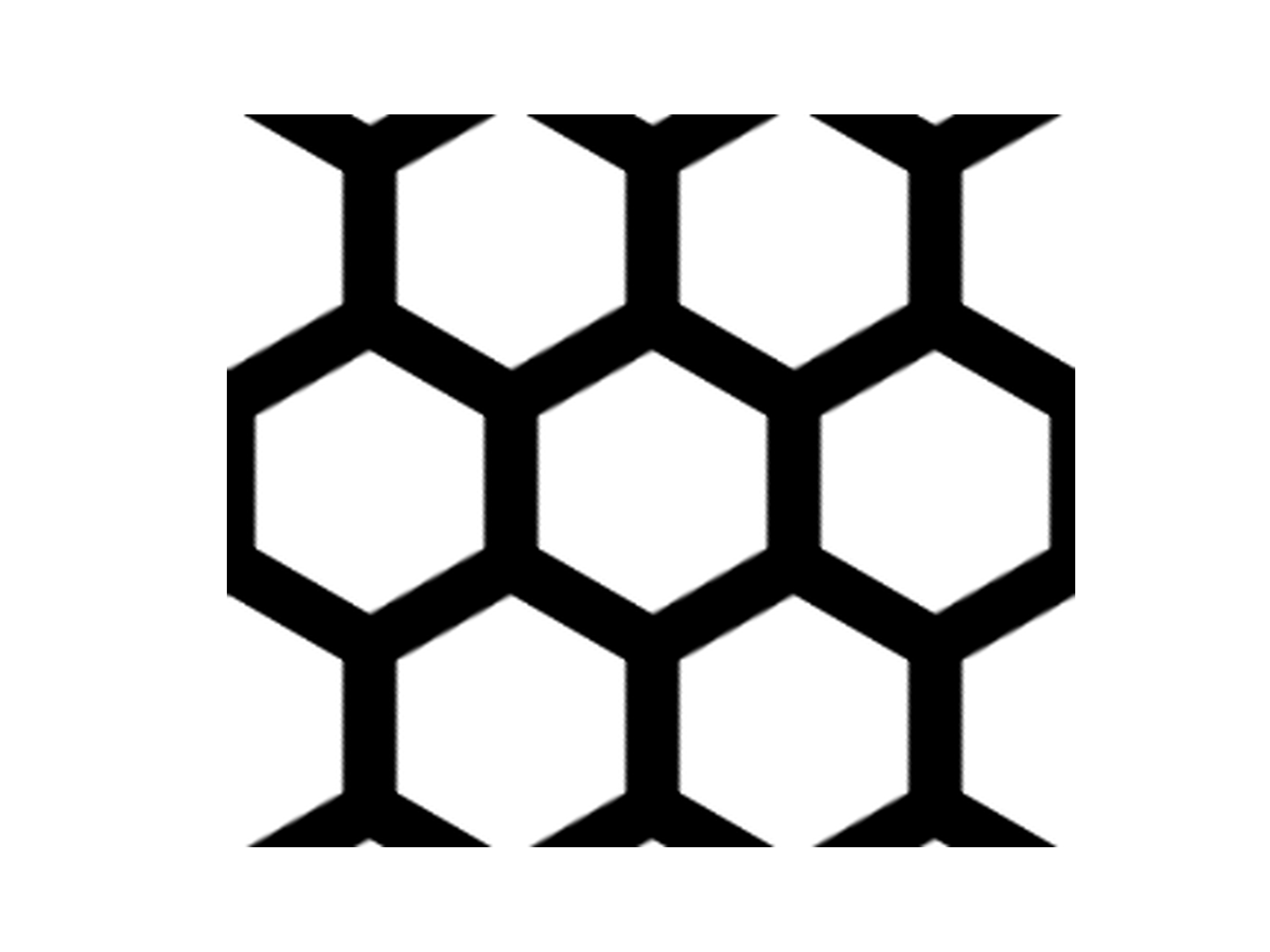}
\end{figure}

On Fig. \ref{fig:hex} we show optimised TE structure (honeycomb network) for triangular geometry for N=1. It corresponds to Fig.\ref{fig:bandgaps}d, N=1 (i.e perfect packing $\hat{OP}_1$). The same structures appears for all "magic band numbers" discussed in the main text. It is easy to realise they all have the same value of bandgap, so in Fig. \ref{fig:bandgaps}d calculations are done only for N=1, and then stars at the "magic band numbers" are set on the same level of $\approx$ 52.5 \%. 
\renewcommand{\arraystretch}{1.2}

\begin{table}
    \centering
    \begin{tabular}{|c|c|c||c|c|}
     \hline
       N  & $d_{symb}$ & d & r & w \\
        \hline
        2 & $ 2^{-1/2}$  & 0.7071 & 0.137 & 0.130 \\
        3 & $ (\sqrt{6}-\sqrt{2})/2$  & 0.5176 & 0.131 & 0.116\\
        4 & $(\sqrt{6}-\sqrt{2})/2$ & 0.5176 & 0.1 & 0.094 \\
        5 & $ 5^{-1/2}$ & 0.4472 & 0.087 & 0.086 \\
        6 & $ (3\sqrt{3}-(6\sqrt{3}+4)^{1/2} +1)/6 $ & 0.4004 & 0.085 & 0.079\\
        7 & $ (\sqrt{3}+1)^{-1} $  & 0.3660 & 0.077 & 0.078 \\
        8 & $ (\sqrt{3}+1)^{-1} $ & 0.3660 & 0.071 & 0.067 \\
        9 & $(5+2\sqrt{3})^{-1/2}$ & 0.3437 & 0.066 & 0.061 \\
        \hline
        10 &  &0.316& 0.061 & 0.059 \\
        11 &  & 0.306  & 0.069  & 0.065 \\
        12 &  & 0.300  & 0.056  & 0.060 \\
        13 &  & 0.278  & 0.059  & 0.055 \\
        14 &  & 0.276  & 0.065 & 0.053 \\
        15 &  & 0.274  & 0.049 & 0.053 \\ 
               \hline             \end{tabular}
    \caption{Properties of optimal packing configurations and optimised photonic crystals in square lattice geometry. The first column: the number N of the generating points in a square torus. The second and the third column: symbolic expression and numerical value of the optimal diameter $d$ (minimal distance between the points in the optimal packing configuration) \cite{Musin12}. The fourth columns gives optimised value $r$ of the radius of rods for maximising bandgap in TM polarisation (corresponding to OP$_1$ solutions in Fig. \ref{fig:bandgaps}a of the main text). The last column gives optimised value $w$ of the width of wires in nets maximising bandgap in TE polarisation (OP$_1$ solutions in Fig. \ref{fig:bandgaps}c of the main text, see also Figs in Supplementory Information)\cite{SuppInf}}
    \label{tab:sqtable}
\end{table}

\begin{table}[hbt]
    \centering
    \begin{tabular}{|c||c|c||c|c|}
     \hline
       N  &  $ \varepsilon_{min} $, TE & w &  $ \varepsilon_{min} $, TM  & r  \\
        \hline
        2  & 4.6 & 0.177 & 3.0 & 0.33 \\
        4  & 2.55 & 0.149 & 2.0 & 0.166 \\
        7  & 2.05 & 0.123 & 1.75 & 0.123 \\
        12 & 2.3 & 0.093  & 2.0 & 0.09 \\
        15 & 2.0 & 0.086 & 1.8 & 0.082 \\
        \hline             \end{tabular}
    \caption{Minimal contrast of optimal structures OP$_1$ for sustaining band gap over Nth band. It is defined merely as minimal value of permittivity $\varepsilon$ of inclusions at which band gaps described in the main text are still non vanishingly small (at least 0.3 \%). For TE modes inclusions are wires with width $w$ and for TM modes they are discs with radius $r$. Host media is the air with permittivity equal to 1. }
    \label{tab:contrast}
\end{table}

\clearpage
\newpage 

\beginsuppinf
\section*{Supplementary Information for "New geometric receipts for design of photonic crystals and metamaterials: optimal toric packings"}

\begin{figure}[hbt]
\caption{Optimal packings of circles in 2D square. From top left to bottom right: N =2,3,4,5. \label{fig:sqpack} }
\centering
\includegraphics[width=0.3\textwidth]{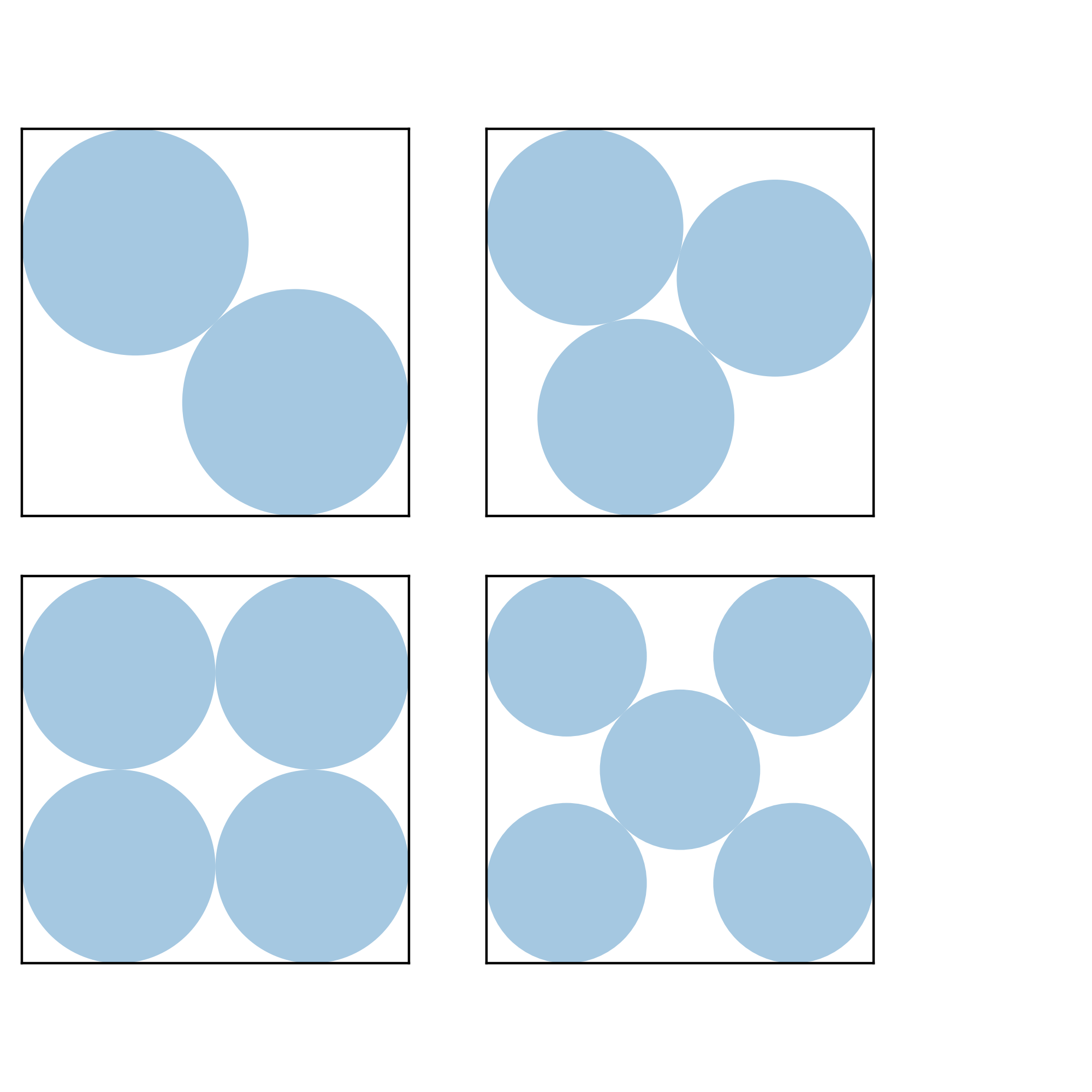}
\end{figure}

In Fig. \ref{fig:sqpack} we show examples of optimal packings in a square, just to clarify its difference from square torus packings.

\begin{figure*}
\caption{ 
Optimised TE structures based on optimal packing (OP$_1$).\label{fig:TEOP1}  }
\centering
\begin{subfigure}[]{0.24\textwidth} 
\subcaption{N=2}
\includegraphics[width=\textwidth]{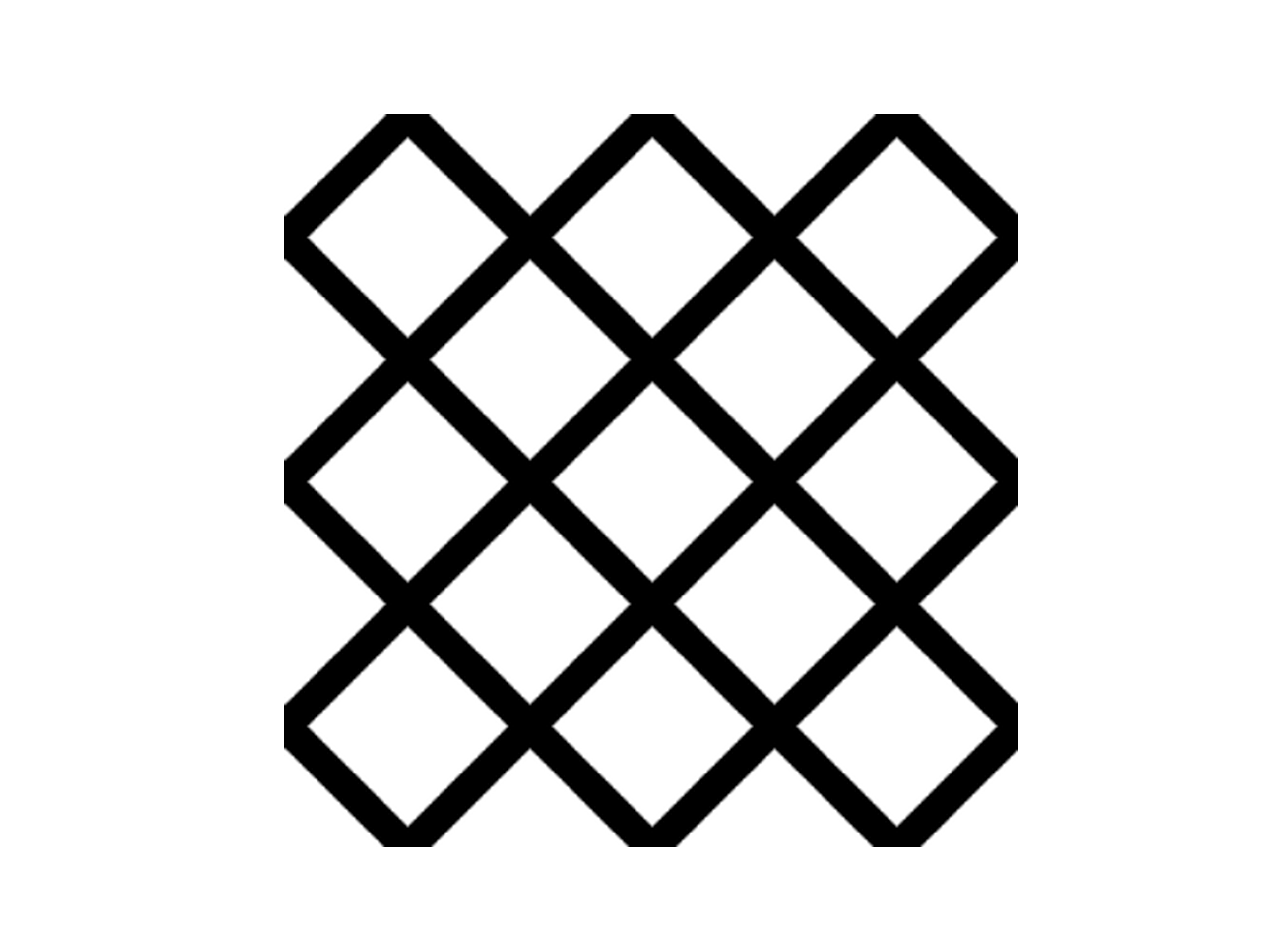}
\end{subfigure}
\begin{subfigure}[]{0.24\textwidth}
\subcaption{N=3}
\includegraphics[width=\textwidth]{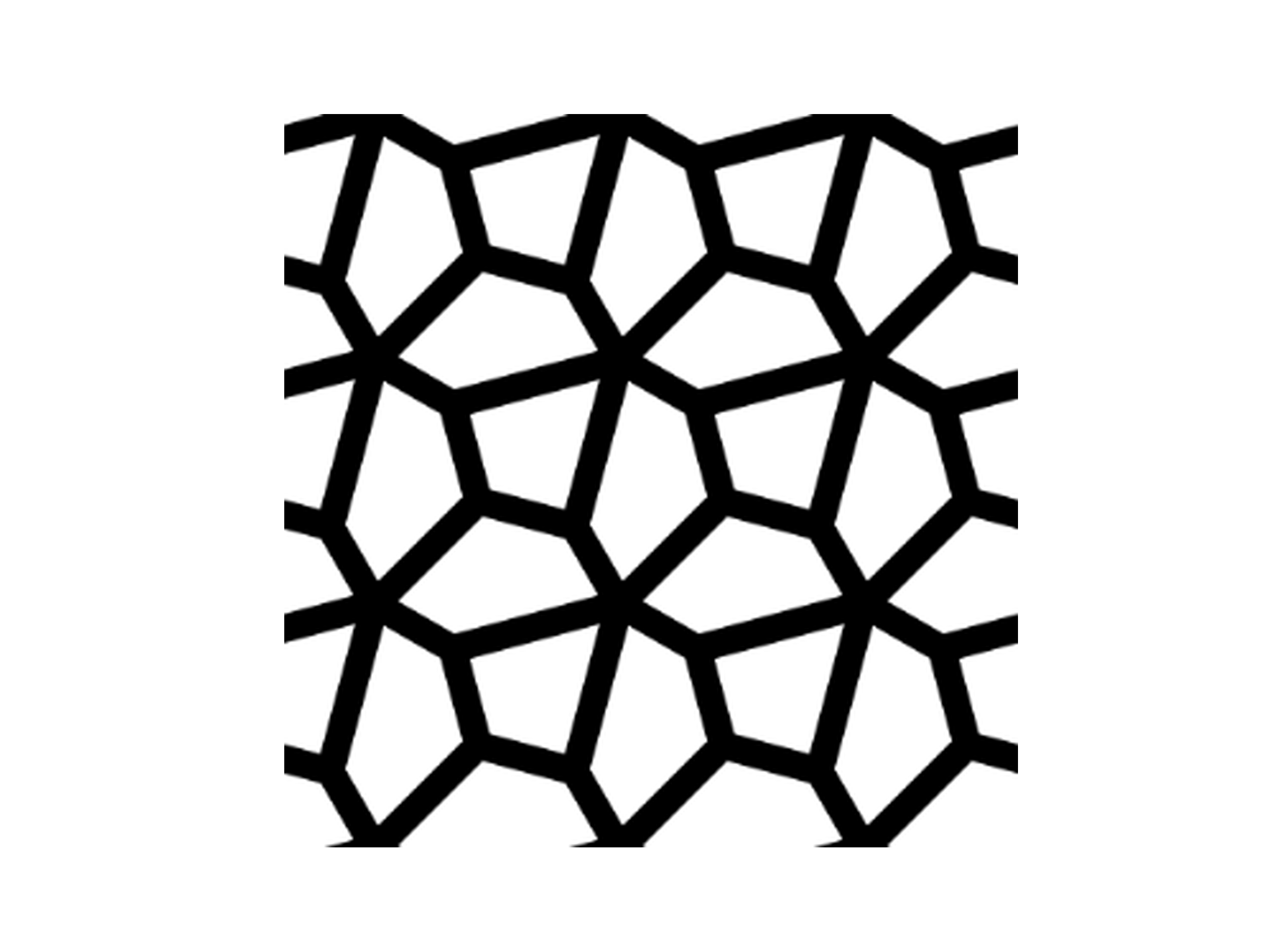}
\end{subfigure}
\begin{subfigure}[]{0.24\textwidth}
\subcaption{N=4}
\includegraphics[width=\textwidth]{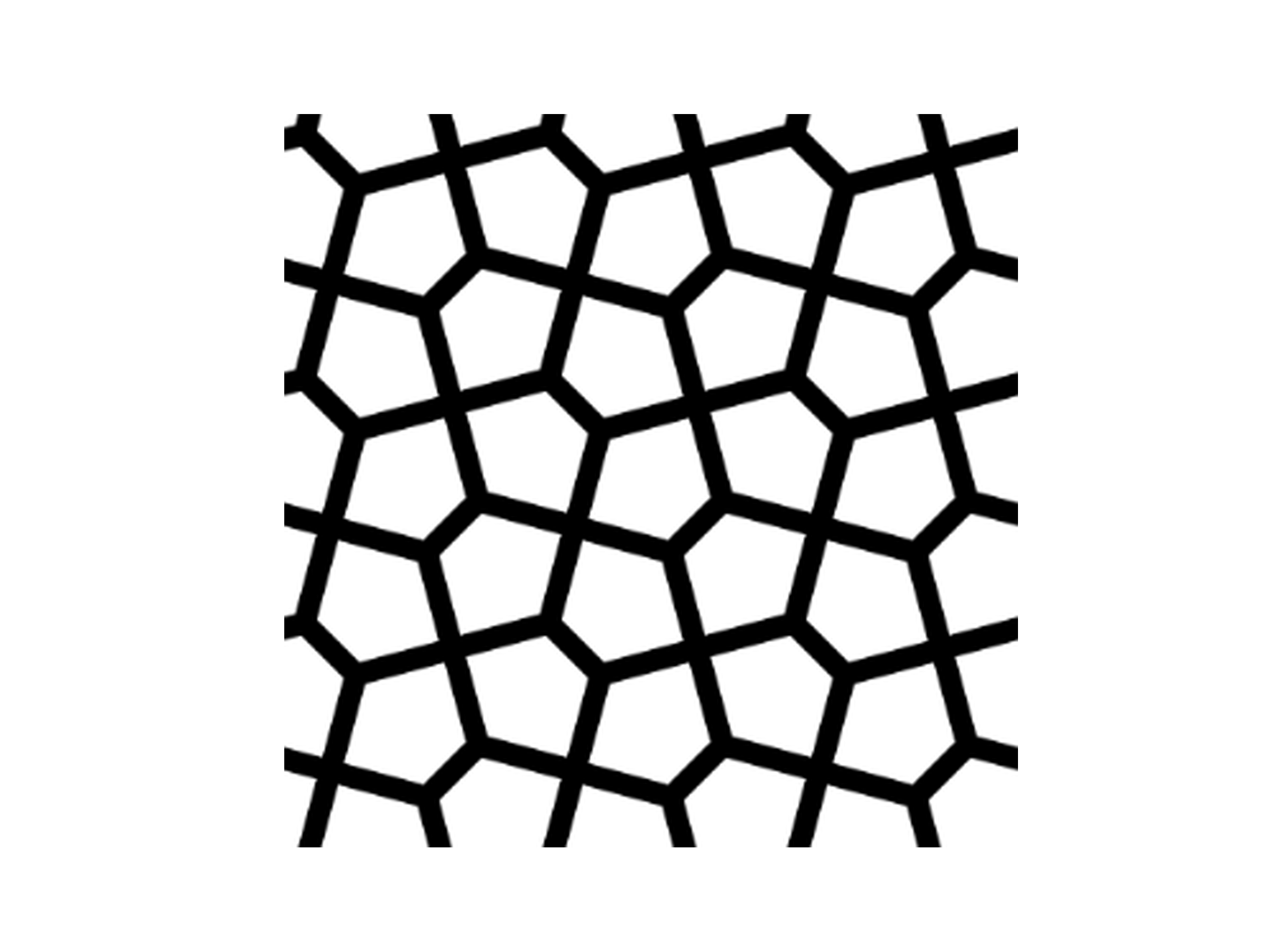}
\end{subfigure}
\begin{subfigure}[]{0.24\textwidth}
\subcaption{N=5}
\includegraphics[width=\textwidth]{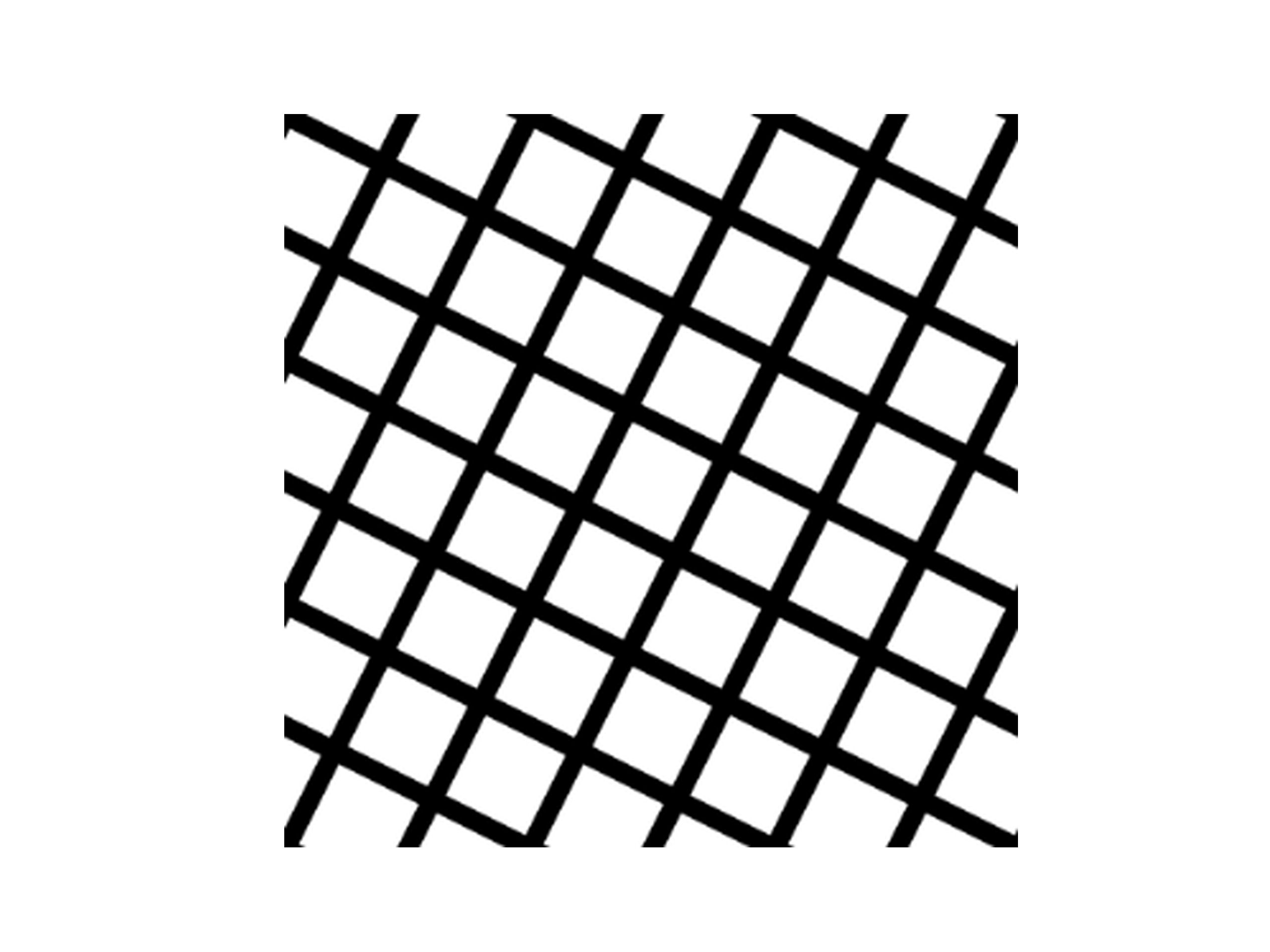}
\end{subfigure}
\begin{subfigure}[]{0.24\textwidth}
\subcaption{N=6}
\includegraphics[width=\textwidth]{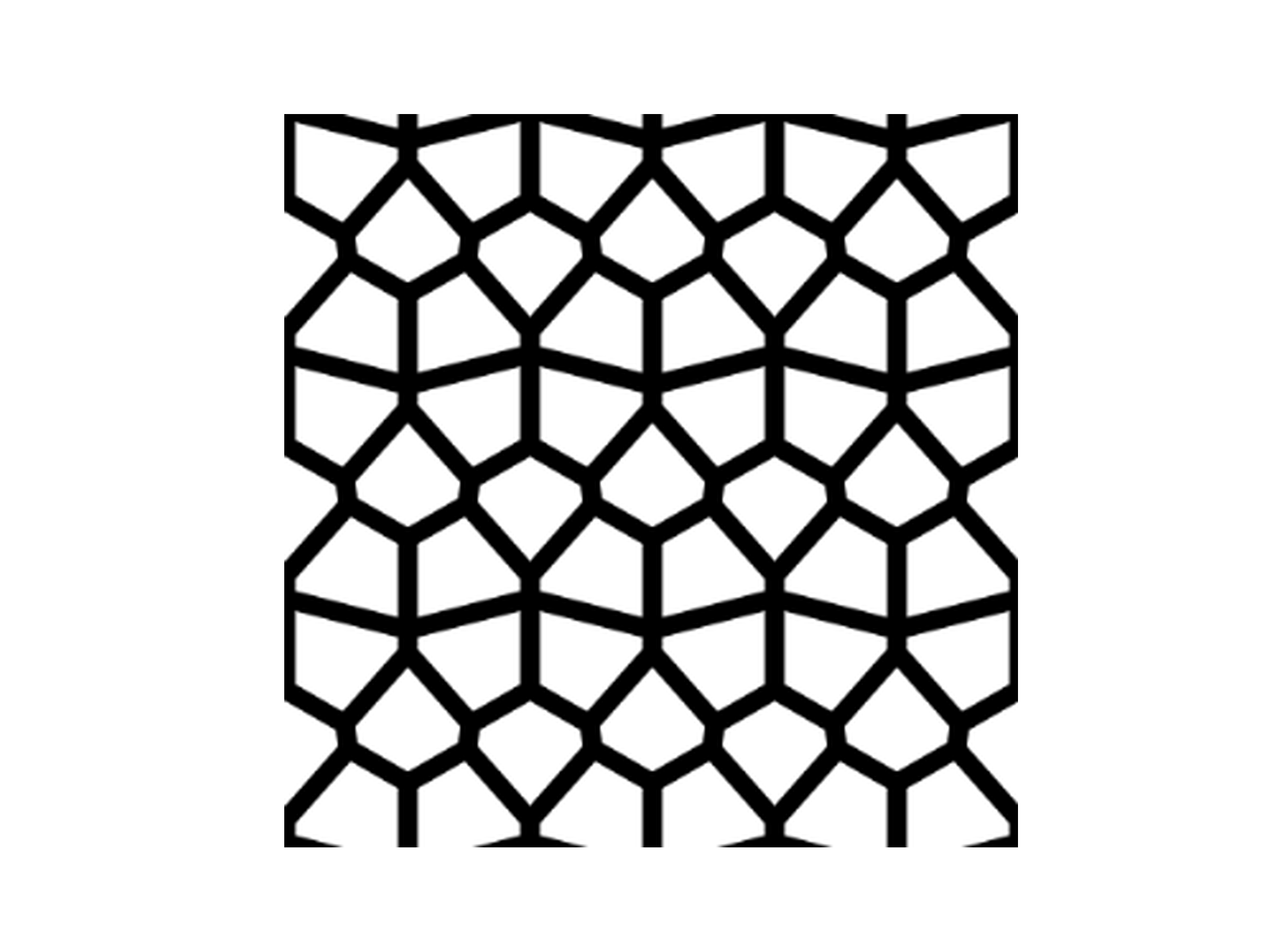}
\end{subfigure}
\begin{subfigure}[]{0.24\textwidth}
\subcaption{N=7}
\includegraphics[width=\textwidth]{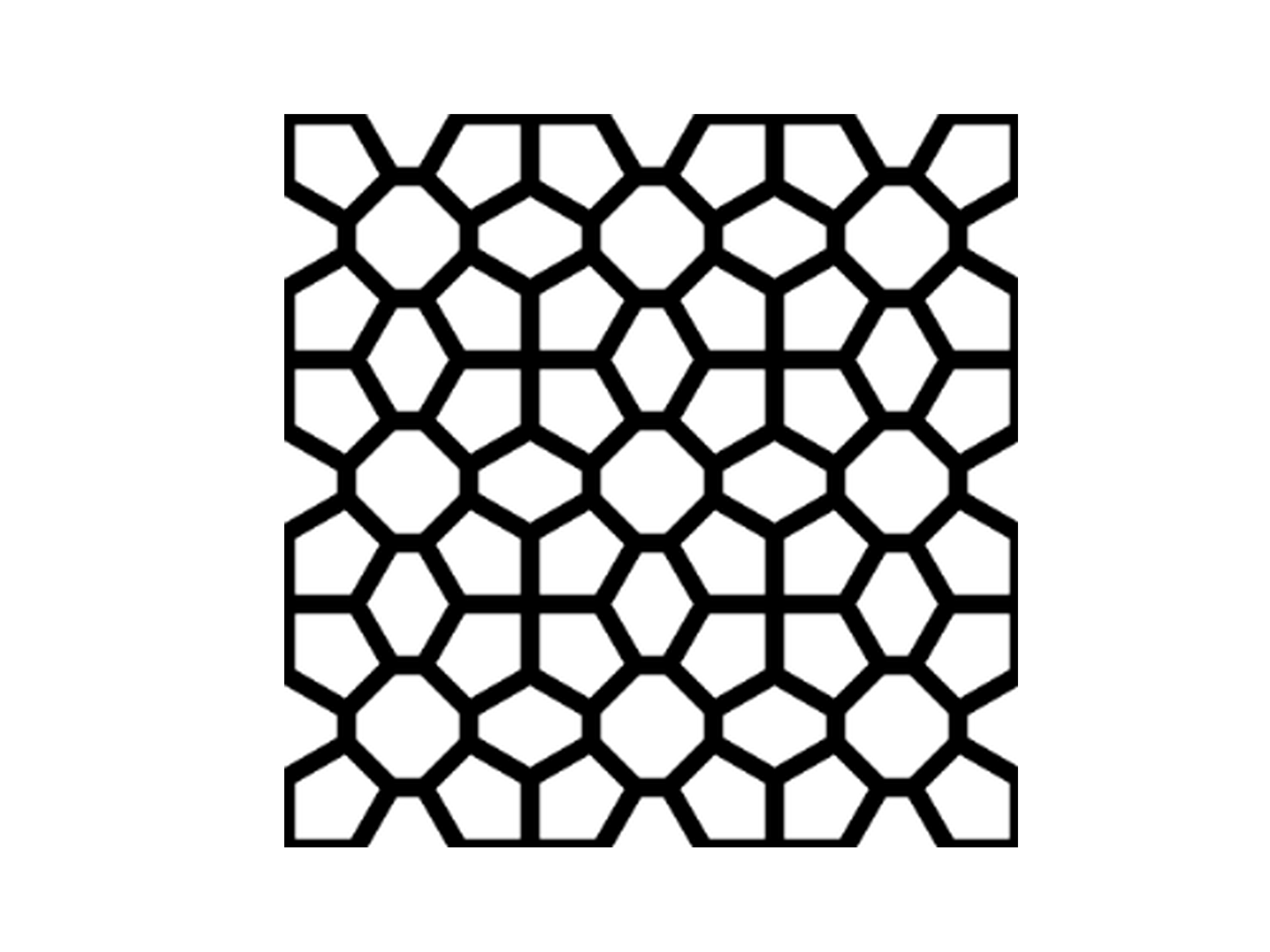}
\end{subfigure}
\begin{subfigure}[]{0.24\textwidth}
\subcaption{N=8}
\includegraphics[width=\textwidth]{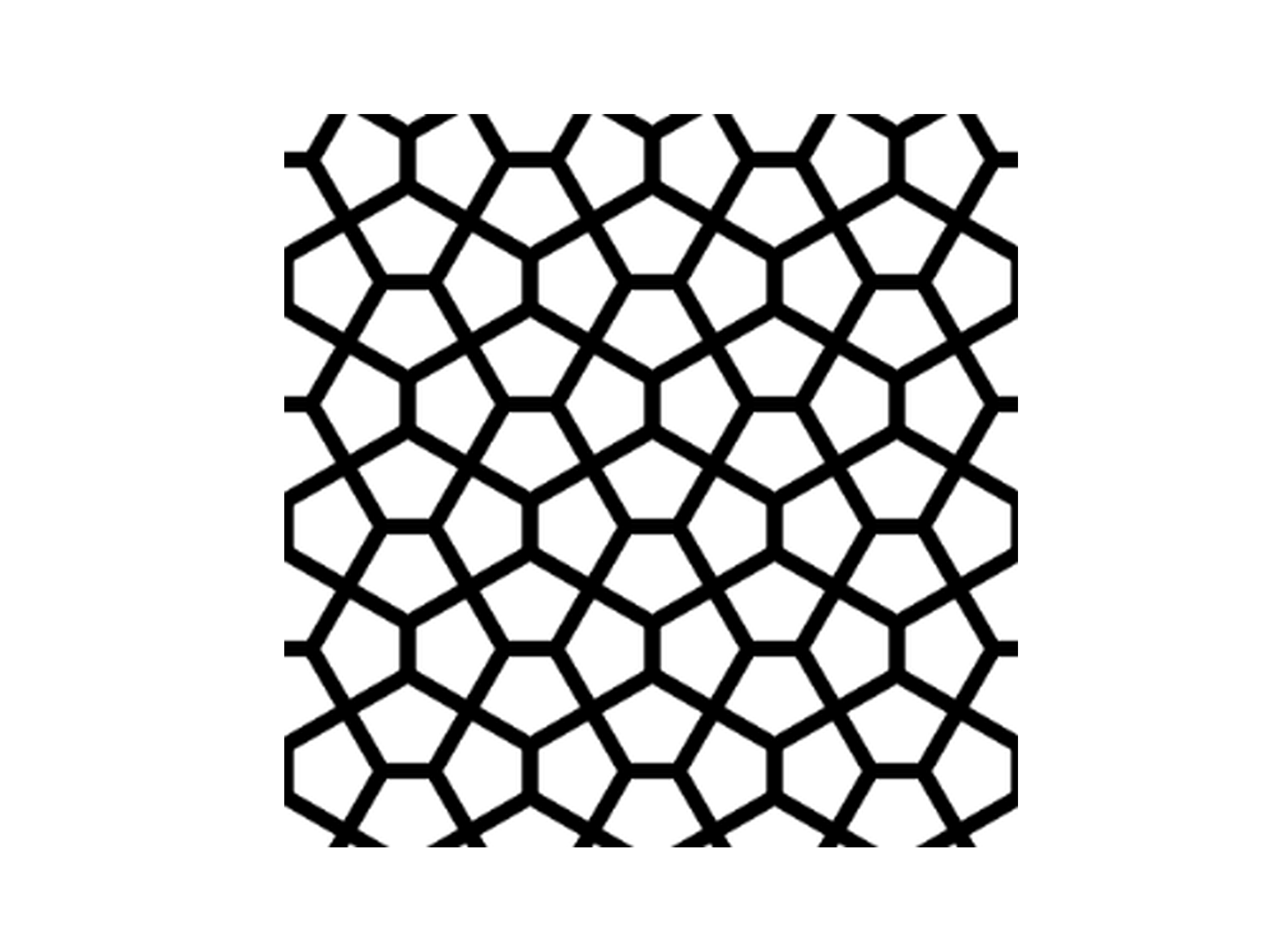}
\end{subfigure}
\begin{subfigure}[]{0.24\textwidth}
\subcaption{N=9}
\includegraphics[width=\textwidth]{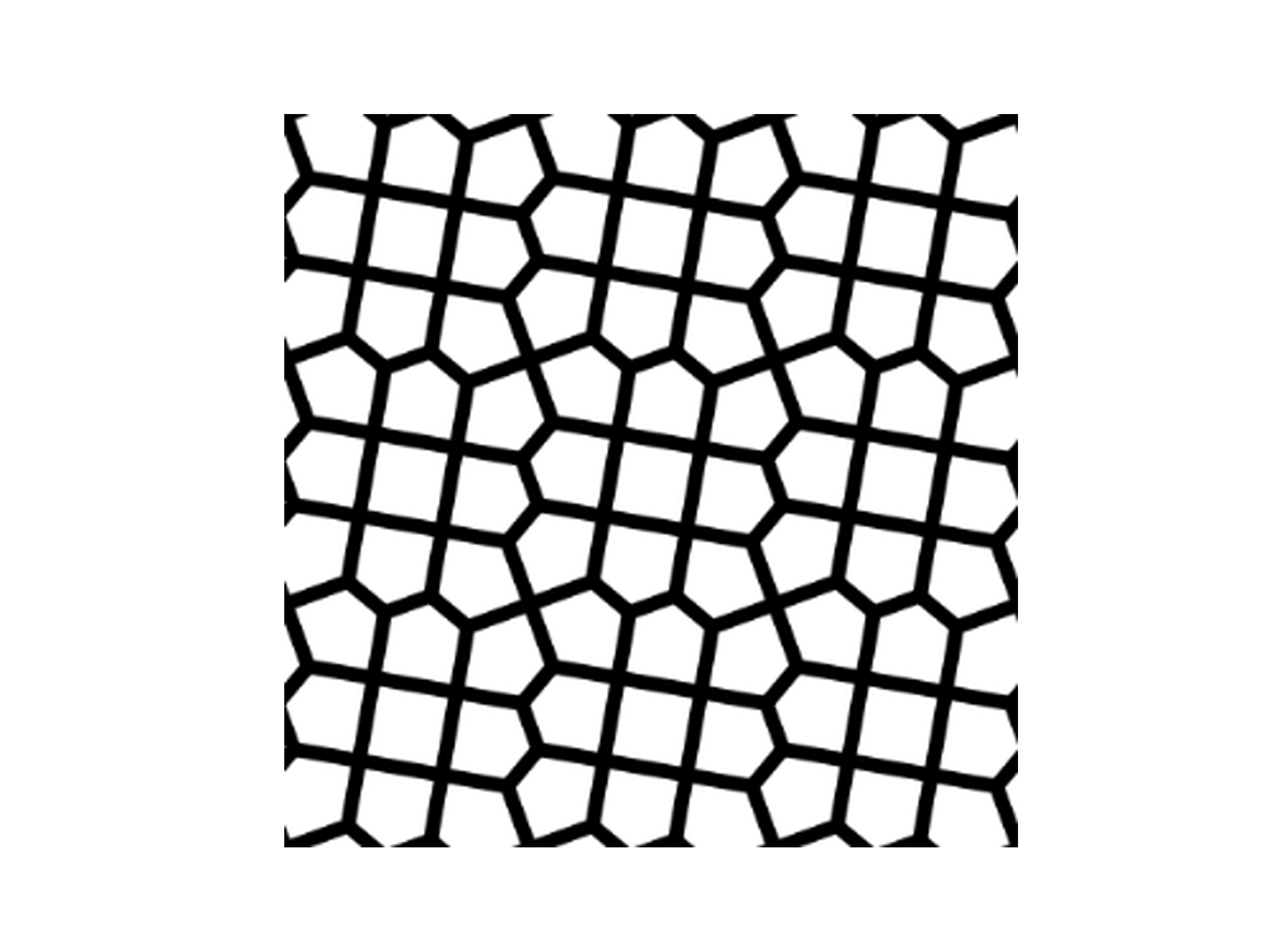}
\end{subfigure}
\end{figure*}

\begin{figure*}[hbt]
\caption{ 
Optimised TE structures based on conjectured optimal packing (OP$_1$), N=10-15.\label{fig:TECOP1}  }

\begin{subfigure}[]{0.26\textwidth}
\subcaption{N=10}
\includegraphics[width=\textwidth]{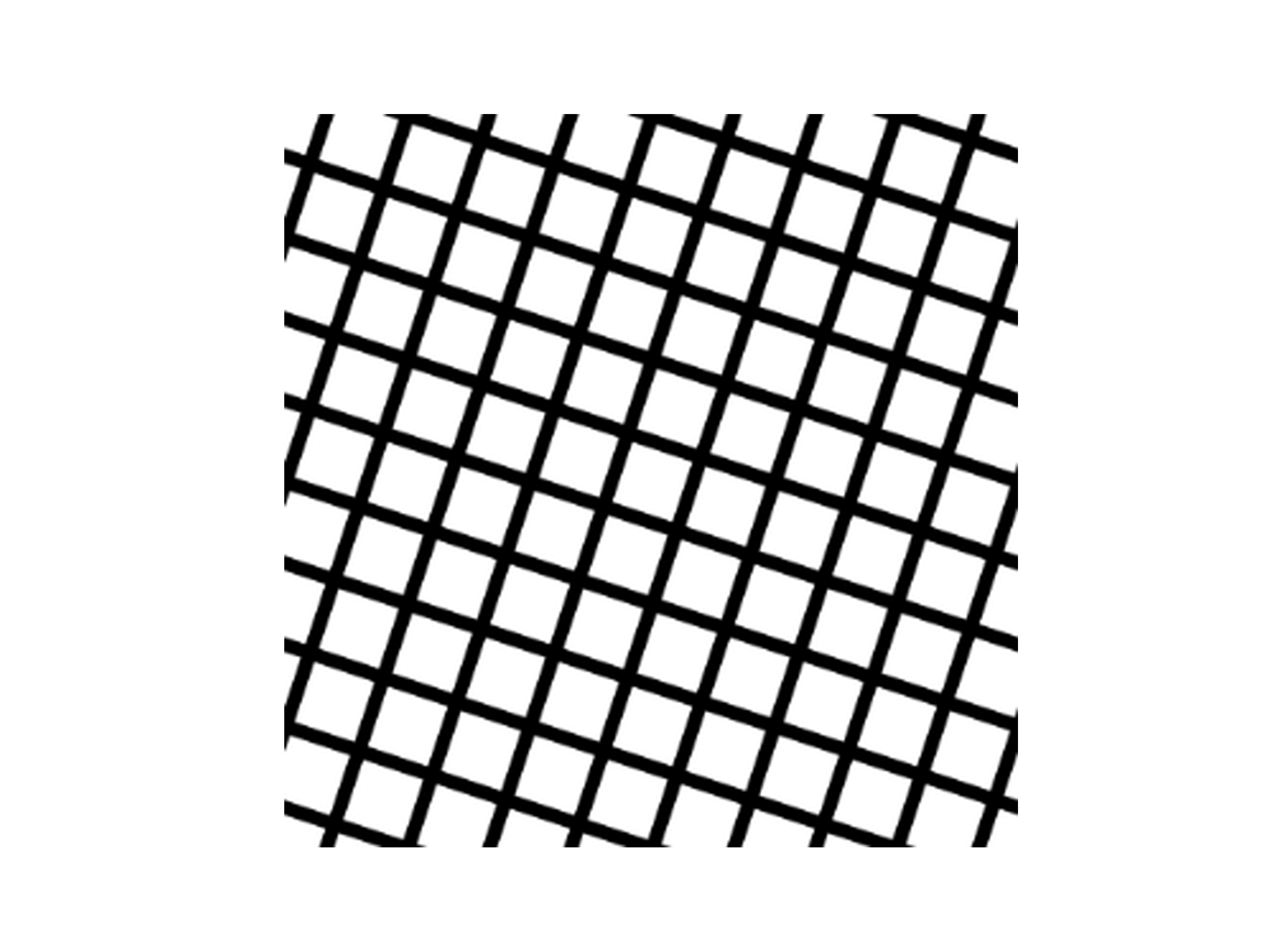}
\end{subfigure}
\begin{subfigure}[]{0.26\textwidth}
\subcaption{N=11}
\includegraphics[width=\textwidth]{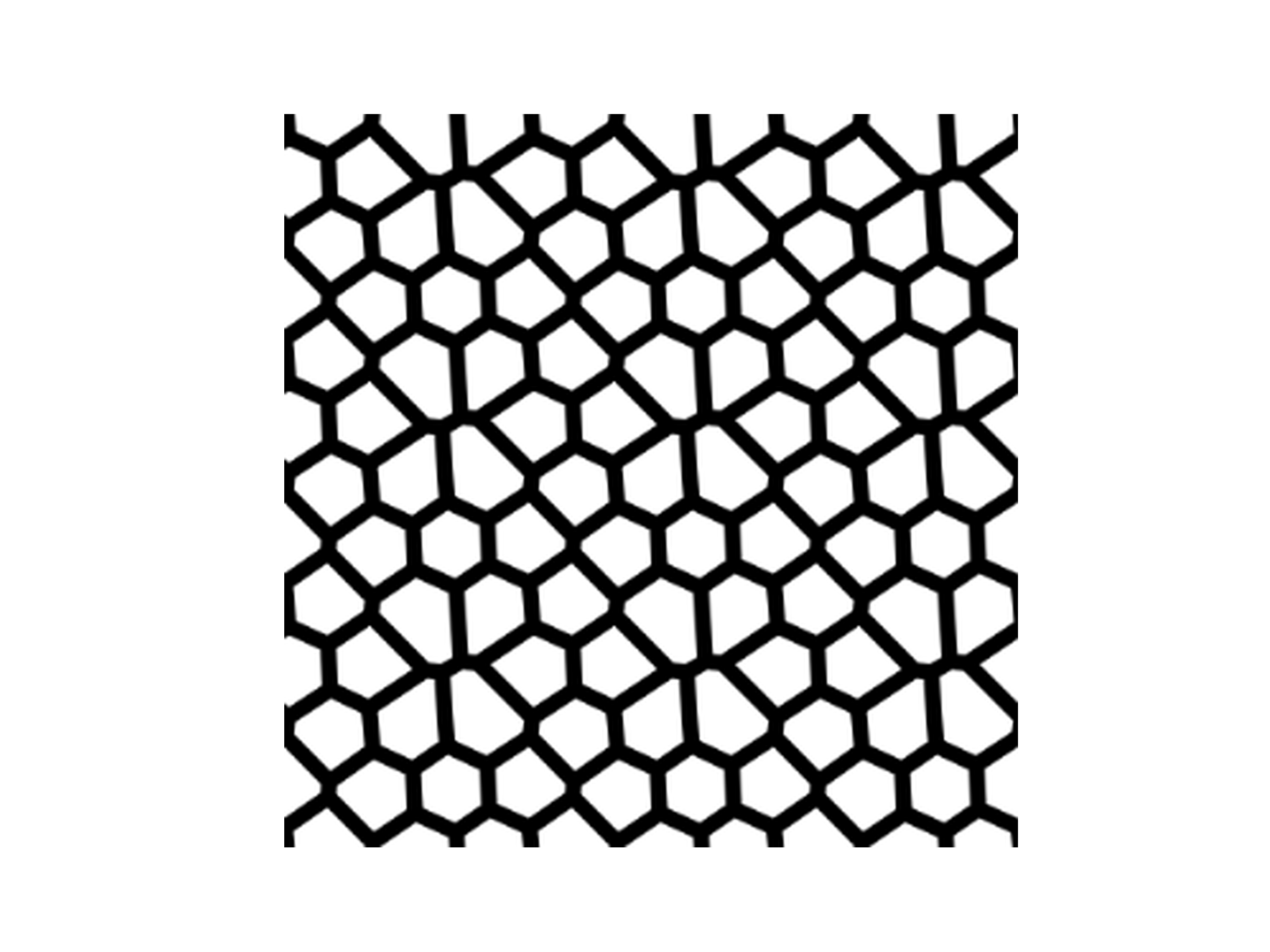}
\end{subfigure}
\begin{subfigure}[]{0.26\textwidth}
\subcaption{N=12}
\includegraphics[width=\textwidth]{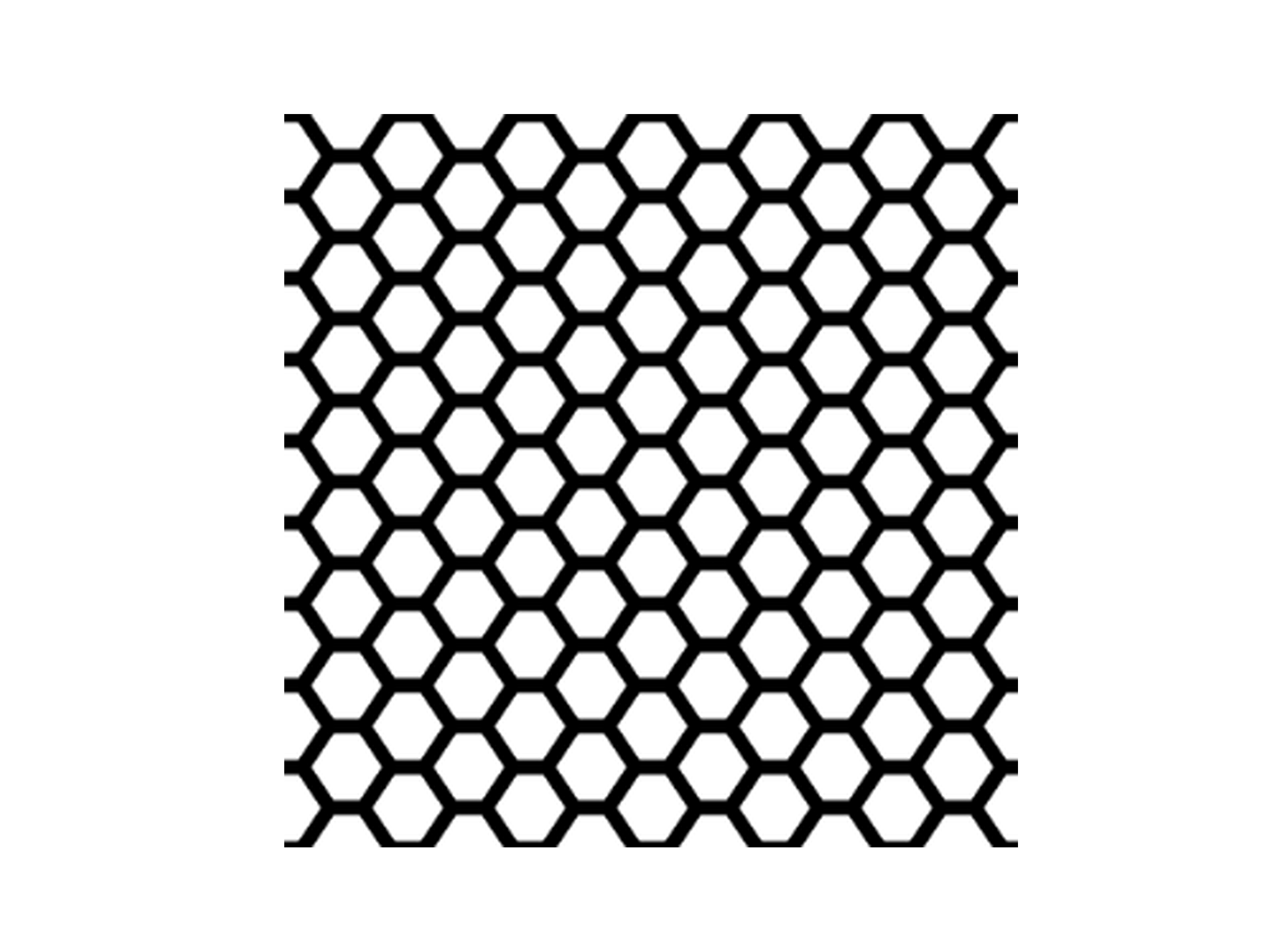}
\end{subfigure}
\begin{subfigure}[]{0.26\textwidth}
\subcaption{N=13}
\includegraphics[width=\textwidth]{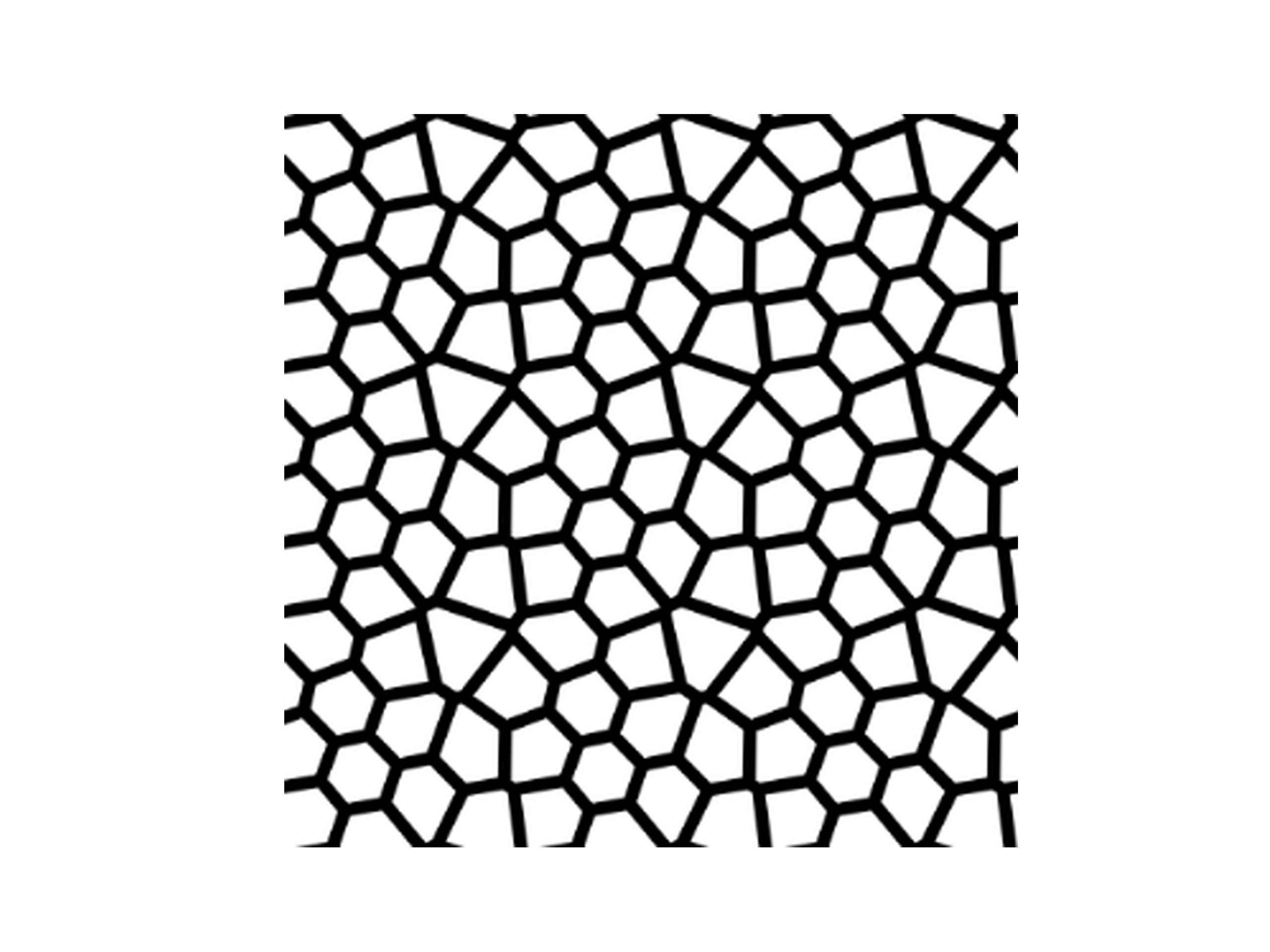}
\end{subfigure}
\begin{subfigure}[]{0.26\textwidth}
\subcaption{N=14}
\includegraphics[width=\textwidth]{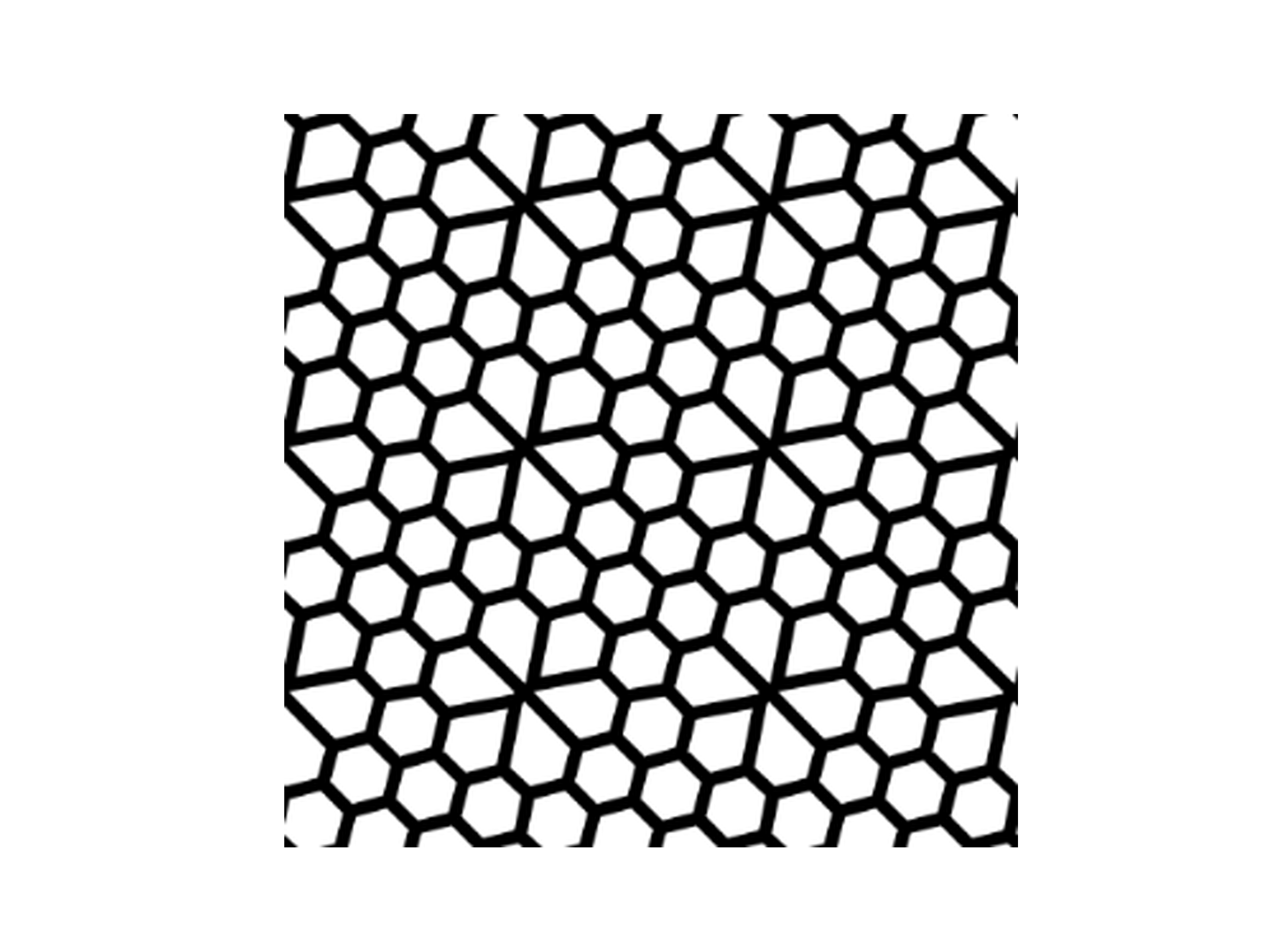}
\end{subfigure}
\begin{subfigure}[]{0.26\textwidth}
\subcaption{N=15}
\includegraphics[width=\textwidth]{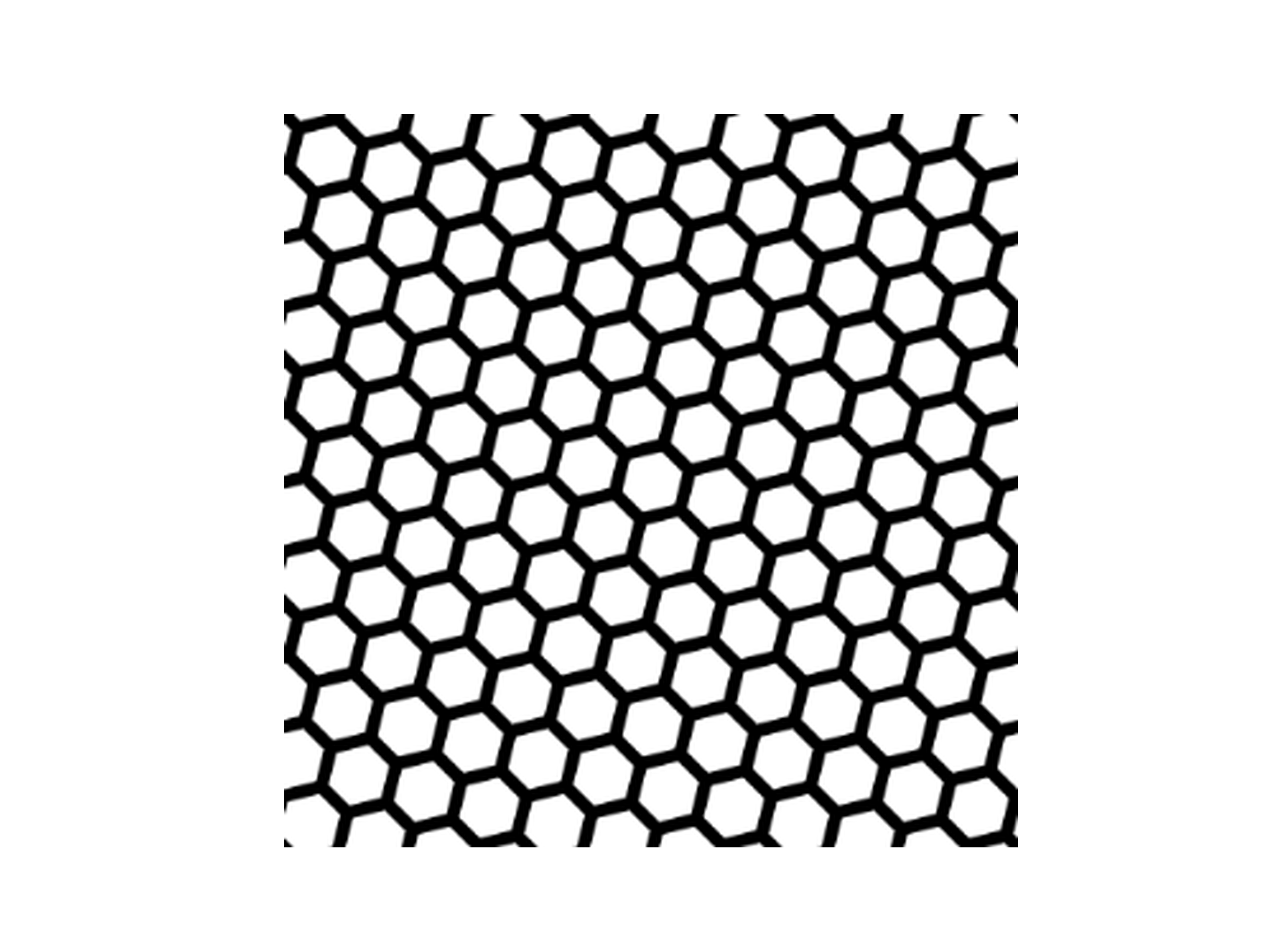}
\end{subfigure}

\end{figure*}

In Figs. \ref{fig:TEOP1} and \ref{fig:TECOP1} we show designs (OP$_1$) optimised for TE modes in square lattice geometry. Optimised values of width $w$ of the nets are given in the Table \ref{tab:sqtable} in Appendix. Fig. \ref{fig:TEOP1} corresponds to N=2 to 9, for which optimal packing configurations are known, and Fig. \ref{fig:TECOP1} to N=10 to 15, for which conjectured optimal packing configurations are known. VT of the optimal or conjectured optimal configurations produce "skeletons" of the nets, and then we vary its width to increase the value of the bandgap.

\end{document}